\documentclass[twocolumn,final,twoside,journal,10pt]{IEEEtran}

\usepackage{epsfig,color}

\usepackage[nolist]{acronym}
\usepackage{graphicx,cite,amsmath,mathrsfs,amsthm}
\usepackage{multirow}
\usepackage{bigstrut}
\usepackage{psfrag}
\usepackage{xcolor}
\usepackage{mathtools}
\mathtoolsset{showonlyrefs}
\usepackage[caption=false]{subfig}
\usepackage{placeins}
\usepackage{relsize}
\usepackage{thmtools}
\usepackage{pgfplots,relsize}
\usepackage[scaled=0.86]{helvet}

\newcommand{\figw}{0.93\columnwidth}

\declaretheoremstyle[spaceabove=6pt, spacebelow=6pt,
headfont=\normalfont\bfseries,
notefont=\normalfont\bfseries, notebraces={(}{)},
bodyfont=\normalfont,
postheadspace=1em]{normalhead}
\declaretheorem[style=normalhead]{step}

\newcommand{\msr}{\mathscr}

\usepackage[varg,cmbraces,uprightGreek]{newtxmath}   
\usepackage{bm}
\renewcommand{\mathbf}{\bm}

\newcommand{\Exp}{\mathsf{E}}
\renewcommand{\t}{{\tiny\mathsf{T}}} 
\renewcommand{\t}{{\mathsf{T}}} 
\newcommand{\rank}[1]{\mathrm{rank}\left[#1\right]}

\newcommand{\reloverhead}{\epsilon}
\newcommand{\absoverhead}{\delta}

\newcommand{\Pf }{\mathsf{P}_{\mathsf{F}}}
\newcommand{\Pfstar }{{\Pf}^\star}
\newcommand{\barPf } { \bar {\mathsf{P}}_{\mathsf{F}}}
\newcommand{\pru}{\mathsf{P}_u}

\newcommand{\osymb}{c} 
\newcommand{\rosymb}{y} 
\newcommand{\rosymbrv}{Y} 
\newcommand{\dmax}{d_{\mathrm{max}}}
\newcommand{\x}{\mathtt{x}}
\renewcommand{\deg}{\mathrm{deg}}
\newcommand{\avgd}{\bar{\Omega}}
\newcommand{\Omegaone}{ \Omega^{(1)} }
\newcommand{\Omegatwox}{ \Omega^{(2)} }
\newcommand{\Omegathree}{ \Omega^{(3)} }
\newcommand{\Omegafour}{ \Omega^{(4)} }
\newcommand{\G}{\mathbf{G}}
\newcommand{\GLTa}{\G}
\newcommand{\Grx}{\mathbf{G}_0}
\newcommand{\Amatrix}{\mathbf{A}}
\renewcommand{\Bmatrix}{\mathbf{B}}
\newcommand{\Cmatrix}{\mathbf{C}}
\newcommand{\Dmatrix}{\mathbf{D}}
\newcommand{\reducedsyst}{\alpha}

\newcommand{\inputset}{\mathscr{V}}
\newcommand{\outputset}{\mathscr{Y}}
\newcommand{\rdeg}{\mathrm{red}}
\newcommand{\rippleset}{\mathscr{R}}
\newcommand{\Ripple}{\mathtt{R}}
\renewcommand{\r}{\mathtt{r}}
\newcommand{\ripple}[1]{ \msr{R}_{#1}}
\newcommand{\ru}{\r_u}
\newcommand{\Ru}{\Ripple_u}

\newcommand{\cloudset}{\mathscr{C}}
\newcommand{\Cloud}{\mathtt{C}}
\renewcommand{\c}{\mathtt{c}}
\newcommand{\cloud}[1]{\msr{C}_{#1}}
\newcommand{\Cu}{\Cloud_u}
\newcommand{\cu}{\c_u}

\renewcommand{\S}[1]{\mathsf{S}_{#1}}
\newcommand{\Erv}{\mathtt{A}}
\newcommand{\erv}{\mathtt{a}}
\renewcommand{\b}{\mathtt{b}}

\newcommand{\bu}{\b_u}

\newcommand{\Y}{{\mathtt{T}}}
\newcommand{\y}{{\mathtt{t}}}
\newcommand{\n}{{\mathtt{t}}}
\newcommand{\Nu}{{\mathtt{T}_u}}
\renewcommand{\nu}{\n_u}

\newcommand{\Raptorinput}{u}
\newcommand{\vecu}{\mathbf{\Raptorinput}}
\newcommand{\Rintermsymbol}{v}
\newcommand{\vecv}{\mathbf{\Rintermsymbol}}
\newcommand{\Rosymb}{c}
\newcommand{\constmatrix}{\mathbf{M}}
\newcommand{\hmatrixpre}{\mathbf{H}_{\mathrm{P}}}
\newcommand{\Rrosymb}{\rosymb}

\newcommand{\weo}{A^{\mathsf o}}
\renewcommand{\l}{l}
\newcommand{\pil}{\pi_{\l}}

\newcommand{\Rijsets}[1]{\mathscr{Z}_{#1}}
\newcommand{\Rijset}[2]{\mathscr{Z}_{#2,#1}}
\newcommand{\Rijcard}[2]{\mathtt{z}_{#2,#1}}
\newcommand{\Rijcards}[1]{\mathtt{z}_{#1}}
\newcommand{\Rijreals}[1]{ \mathtt{Z}_{#1}}
\newcommand{\Rijreal}[2]{ \mathtt{Z}_{#2,#1}}

\newcommand{\rij}[2]{\lambda_{#2,#1}}
\newcommand{\probtx}[2]{\mathsf{P}_{#2,#1}}
\newcommand{\ntx}[2]{B_{#2,#1}}

\newcommand{\Rijcardv}[1]{ \pmb{\mathtt{z}}_{#1}}

\newcommand{\zeros}{\mathbf{0}}

\newcommand{\f}{\mathtt{f}}
\newcommand{\F}{\mathtt{F}}

\newcommand{\eventa}{\rosymbrv \in \ripple{u-1}}
\newcommand{\eventb}{ \rosymbrv \in \cloud{u}}

\newtheorem{mydef}{Definition}
\newtheorem{prop}{Proposition}
\newtheorem{theorem}{Theorem}

\newtheorem{example}{Example}

\begin{acronym}
	\acro{ARQ}{automatic retransmission query}
	\acro{AWGN}{additive white Gaussian noise}
	\acro{BCH}{Bose Chaudhuri Hocquenghem}
	\acro{BEC}{binary erasure channel}
	\acro{BSC}{binary symmetric channel}
	\acro{BP}{belief propagation}
	\acro{CER}{codeword error rate}
	\acro{C-OWEF}{conditional output-weight enumerator function}
	\acro{CRC}{cyclic redundancy check}
	\acro{FEC}{forward error correction}
	\acro{GE}{Gaussian elimination}
	\acro{GRS}{generalized Reed-Solomon}
	\acro{HDPC}{high-density parity-check}
	\acro{ID}{inactivation Decoding}
	\acro{i.i.d.}{independent and identically distributed}
	\acro{IOWE}{input output weight enumerator}
	\acro{MDS}{maximum distance separable}
	\acro{ML}{maximum likelihood}
	\acro{MWI}{maximum weight inactivation}
	\acro{LDGM}{low-density generator matrix}
	\acro{LDPC}{low-density parity-check}
    \acro{LRFC}{linear random fountain code}
	\acro{LT}{Luby transform}
	\acro{PMF}{probability mass function}
	\acro{QEC}{$q$-ary erasure channel}
	\acro{RI}{random inactivation}
	\acro{RS}{Reed-Solomon}
	\acro{RSD}{robust soliton distribution}
	\acro{SA}{simulated annealing}
	\acro{SPC}{single parity-check}
	\acro{TCP}{Transmission Control Protocol}
	\acro{WE}{weight-enumerator}
	\acro{WEF}{weight-enumerator function}
\end{acronym}

\begin{document}


\title{Inactivation Decoding of LT and Raptor Codes: Analysis and Code Design}

\author{
	Francisco L\'azaro, \IEEEmembership{Student Member, IEEE}, Gianluigi Liva, \IEEEmembership{Senior Member, IEEE},\\
	Gerhard Bauch, \IEEEmembership{Fellow, IEEE}
	\thanks{Francisco L\'azaro and Gianluigi Liva are with the Institute of Communications and
		Navigation of the German Aerospace Center (DLR), Muenchner Strasse 20, 82234
		Wessling, Germany.
		Email:\{\texttt{Francisco.LazaroBlasco}, \texttt{Gianluigi.Liva}\}\texttt{@dlr.de}.}
	\thanks{Gerhard Bauch is with the Institute for Telecommunication,  Hamburg University of Technology, Hamburg, Germany.
		E-mail: \texttt{Bauch@tuhh.de}.}
	\thanks{Corresponding Address:
		Francisco L\'azaro, KN-SAN, DLR, Muenchner Strasse 20, 82234 Wessling, Germany. Tel: +49-8153 28-3211, Fax: +49-8153 28-2844, E-mail: \texttt{Francisco.LazaroBlasco@dlr.de}.}
	\thanks{This work has been presented in part at the 54th Annual Allerton Conference on Communication, Control, and Computing, Monticello, Illinois, USA, September 2015 \cite{lazaro:Allerton2015}, and at the 2014 IEEE Information Theory Workshop, Hobart, Tasmania, November 2014 \cite{lazaro:ITW}.
	}
\thanks{This work has been accepted for publication in IEEE Transactions on Communications, DOI: 10.1109/TCOMM.2017.2715805 }
\thanks{\copyright 2017 IEEE. Personal use of this material is permitted. Permission
from IEEE must be obtained for all other uses, in any current or future media, including
reprinting /republishing this material for advertising or promotional purposes, creating new
collective works, for resale or redistribution to servers or lists, or reuse of any copyrighted
component of this work in other works}
} \maketitle

\thispagestyle{empty}


\markboth{submitted to IEEE Transactions on Communications}{Francisco L\'azaro et al.: Inactivation Decoding of LT and Raptor Codes: Analysis and Code Design} %

\begin{abstract}
	In this paper we analyze LT and Raptor codes under inactivation decoding. A first order analysis is  introduced, which provides the expected number of inactivations for an LT code, as a function of the output distribution, the number of input symbols and the decoding overhead. The analysis is then extended to the calculation of the distribution of the number of  inactivations. In both cases, random inactivation is assumed. The developed analytical tools are then exploited to design LT and Raptor codes, enabling a tight control on the decoding complexity vs. failure probability trade-off. The accuracy of the approach is confirmed by numerical simulations.
\end{abstract}





\section{Introduction}

\IEEEPARstart{F}{ountain} codes \cite{Metzner84,byers02:fountain} provide  an efficient solution for data delivery to large user populations over broadcast channels. The result is attained by encoding the data object (e.g., a file) through an $(n,k)$ linear code, where the number of \emph{output} (i.e., encoded) symbols $n$ can grow indefinitely, enabling a simple rate adaptation to the channel conditions. Due to this, fountain codes are often regarded as \emph{rateless} codes.
Fountain codes were originally conceived for transmission over erasure channels. In practice, different users experience a different channel quality resulting in a different erasure probability. When an efficient fountain code is employed, after a user has received $m=k+\absoverhead$ output symbols, with $\absoverhead$ small, the original input symbols can be recovered with high probability.

\ac{LT} codes were introduced in \cite{luby02:LT} as a first example of practical fountain codes. In \cite{luby02:LT} an iterative erasure decoding algorithm was introduced that performs remarkably well when the number of input symbols $k$ is large. Raptor codes  \cite{maymounkov2002online,shokrollahi04:raptor,shokrollahi06:raptor} address some of the shortcomings of \ac{LT} codes. A Raptor code is a serial concatenation of an $(h,k)$ outer linear block code with an inner \ac{LT} code.

Most of the literature on \ac{LT} and Raptor codes considers iterative decoding  (see e.g. \cite{Karp2004,maneva2006new,Puducheri07,Hyytia07,Vukobratovic10,Maatouk:2012,Shirvanimoghaddam13}).
In \cite{Karp2004,shokrollahi2009theoryraptor,Maatouk:2012} \ac{LT} codes under iterative decoding were analyzed using a dynamic programming approach. This analysis models the iterative decoder as a finite state machine and it can be used to derive the probability of decoding failure of \ac{LT} codes under iterative decoding.
Iterative decoding is particularly effective for large input block sizes, with $k$ at least in the order of several tens of thousands symbols \cite{shokrollahi06:raptor}.
However, in practice, moderate and small values of $k$ are often used, due to memory limitations at the receiver side, or due to the fact that the piece of data to be transmitted is small. For example, the recommended value of $k$ for the Raptor codes standardized in \cite{MBMS12:raptor} is between $1024$ and $8192$ symbols. In this regime, iterative decoding of \ac{LT} and Raptor codes turns to be largely sub-optimum. In actual implementations, different decoding algorithms are used. In particular, an efficient \ac{ML} decoding algorithm usually referred to as \emph{inactivation decoding} \cite{shokrollahi2005systems,shokrollahi2011raptor} is frequently employed.

Inactivation decoding belongs to a large class of Gaussian elimination algorithms tailored to the solution of large sparse linear systems \cite{studio3:lanczos52,berlekamp68:algebraic,lamacchia91:solving,Richardson01,miller04:bec,paolini2012maximum}. The algorithm can be seen as an extension of iterative decoding, where whenever the iterative decoding process stops, a variable (input symbol) is declared as \emph{inactive} (i.e., removed from the equation system) and iterative decoding is resumed. At the end of the process, the inactive variables have to be solved using Gaussian elimination. If a unique solution is found, it is possible to recover all the remaining input symbols by back-substitution (i.e., using iterative decoding).
A few works addressed the performance of \ac{LT} and Raptor codes under inactivation decoding. The decoding failure probability of several types of \ac{LT} codes was thoroughly analyzed via tight lower and upper bounds in \cite{schotsch2011performance,Vary2011:Allerton,schotsch:2013,Schotsch:14}. In \cite{lazaro:JSAC} the distance spectrum of \emph{fixed-rate} Raptor codes is characterized, which enables the evaluation of their performance under \ac{ML} erasure decoding (e.g., via upper union bounds \cite{CDi2001:Finite,Liva2013}). Upper bounds on the failure probability for binary and non-binary Raptor codes are introduced in \cite{lazaro:Globecom2016}. However, none of these works addressed inactivation decoding from a complexity viewpoint. In fact, the complexity of inactivation decoding grows with the  third power of the number of inactivations \cite{miller04:bec}. It is hence of large practical interest to develop a code design which leads (on average) to a small number of inactivations.

In \cite{paolini2012maximum}, inactivation decoding of \ac{LDPC} codes was analyzed, providing a reliable estimate of the expected number of inactivations. The approach of \cite{paolini2012maximum} relies on bridging the inactivation decoder with the Maxwell decoder of \cite{Measson:BridgeBPML}. By establishing an equivalence between inactive and guessed variables, it was shown how to exploit the Area Theorem \cite{Ashikhmin:AreaTheorem} to derive the average number of inactivations. Though, the approach of \cite{paolini2012maximum} cannot be extended to \ac{LT} and Raptor codes due to their rateless nature. In \cite{mahdaviani2012raptor} the authors proposed a new degree distribution design criterion to design the \ac{LT} component of Raptor codes assuming inactivation decoding. However, when this criterion is employed there is not a direct control on the average output degree.	
In \cite{Chingbats}, the authors present a finite length analysis of batched sparse codes under inactivation decoding that provides the expected number of inactivations needed for decoding.

In this paper we provide a thorough analysis of LT and Raptor codes under inactivation decoding. A first order analysis is  introduced, which allows estimating the expected number of inactivations for an LT code, as a function of the output distribution, the number of input symbols and the decoding overhead.
The analysis is then extended to the calculation of the distribution of the number of inactivations. In both cases, random inactivation is assumed. The developed analytical tools are then exploited to design LT and Raptor codes, enabling a tight control on the decoding complexity vs. failure probability trade-off. The work presented in this paper is an extension of the work in \cite{lazaro:Allerton2015}. In particular, the different proofs in \cite{lazaro:Allerton2015} have been modified and extended. Furthermore, in this work we not only consider LT codes but also a class of Raptor Codes.
An analysis based on a Poisson assumption approximation is presented in the Appendix\footnote{{In the Appendix we provide a simplified analysis, which,  though related to the one in \cite{lazaro:ITW}, is fully developed here on the basis of the analysis provided in Section~\ref{chap:inact_analysis} of this paper. The simplified analysis is based on a Poisson approximation, which is analyzed in depth in the Appendix, discussing its limitations and comparing the analytical results with Monte Carlo simulations.}}.

The rest of the paper is structured as follows. In Section~\ref{sec:sys} we present the system model considered in this paper. Section~\ref{chap:LT_inact} contains a detailed description of inactivation decoding applied to \ac{LT} codes. In Section~\ref{chap:inact_analysis} we focus on the analysis of \ac{LT} codes under inactivation decoding. Section~\ref{chap:raptor_inactivation_decoding} shows how to exploit the analysis of the LT component of a Raptor code to estimate the number of inactivations for the Raptor code. A Raptor code design methodology is presented too. The conclusions are presented in Section~\ref{sec:conclusion}.

\section{Preliminaries}\label{sec:sys}
Let $\vecv = (v_1, v_2, ...,v_k)$ be the $k$ input symbols, out of which the \ac{LT} encoder generates the output symbols ${\mathbf{\osymb} = (\osymb_1, \osymb_2, ...,\osymb_n)}$, where $n$  can grown indefinitely. Each output symbol is generated by first sampling a degree $d$ from an output degree distribution ${\Omega= (\Omega_1,\Omega_2,\Omega_3,\hdots\Omega_{\dmax})}$, where $\dmax$ is the maximum output degree. In the following, we will make use of a polynomial representation of the degree distribution in the form
\[
\Omega(\x):=\sum_d \Omega_d \x^d
\]
where $\x$ is a dummy variable.
Then, $d$ distinct input symbols are selected uniformly at random and their x-or is computed to generate the output symbol. The relation between output and input symbols can be expressed as
\[
\mathbf{\osymb} = \mathbf{v} \GLTa
\]
where $\GLTa$ is the generator matrix of the \ac{LT} code.

The output symbols are transmitted over a \ac{BEC} at the output of which the receiver collects $m=k+\absoverhead$ output symbols, where $\absoverhead$ is referred to as absolute receiver overhead. Similarly, the relative receiver overhead is defined as $\reloverhead:= \delta / k$.
Denoting by $\mathbf{\rosymb}=(\rosymb_1, \rosymb_2, \ldots, \rosymb_m)$ the $m$ received output symbols and  by $\mathcal{I} = \{i_1, i_2, \hdots, i_m \}$ the set of indices corresponding to the $m$ non-erased symbols, we have
\[
\rosymb_j = \osymb_{i_j}.
\]
This allows us to express the dependence of the received output symbols on  the input symbols as
\begin{align}
	\Grx^\t \mathbf{v}^\t= \mathbf{\rosymb}^\t
	\label{eq:ml_eq_sys}
\end{align}
with $\Grx$ given by the $m$ columns of $\GLTa$ with indices in $\mathcal{I}$.
\ac{ML} erasure decoding requires the solution of \eqref{eq:ml_eq_sys}. Note that decoding is successful (i.e., the unique solution can be found) if and only if $\rank{\Grx} =k$.

\section{Inactivation Decoding of LT Codes}\label{chap:LT_inact}

\begin{figure}
	\centering
	\includegraphics[width=0.6\columnwidth]{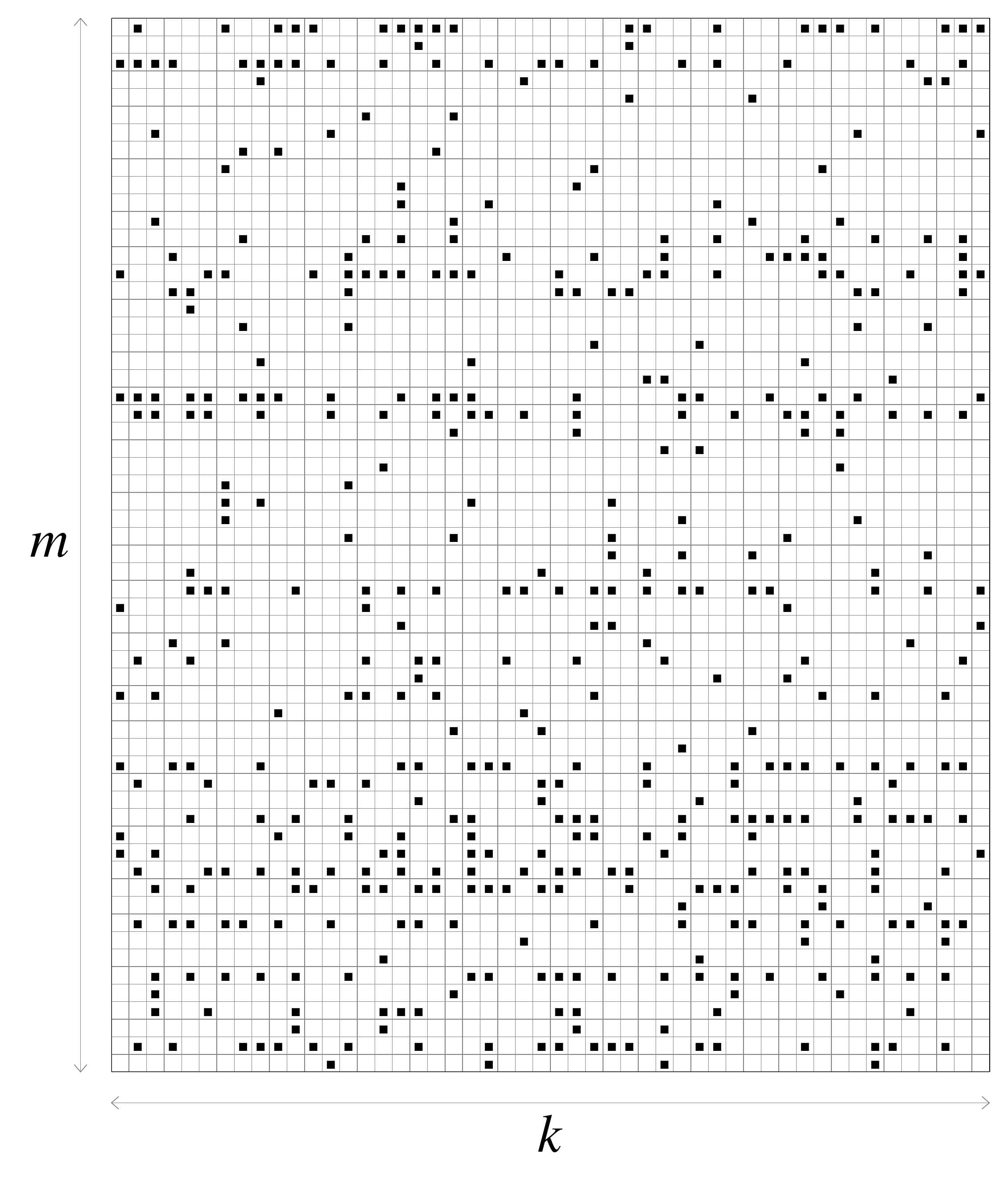}
	\caption{Structure of the matrix $\Grx^\t$.}\label{fig:inactivation_0}
\end{figure}

\begin{figure*}
	\centering
	\subfloat[Triangulation process.\label{fig:inactivation_1}]{\centering\includegraphics[height=0.55\columnwidth]{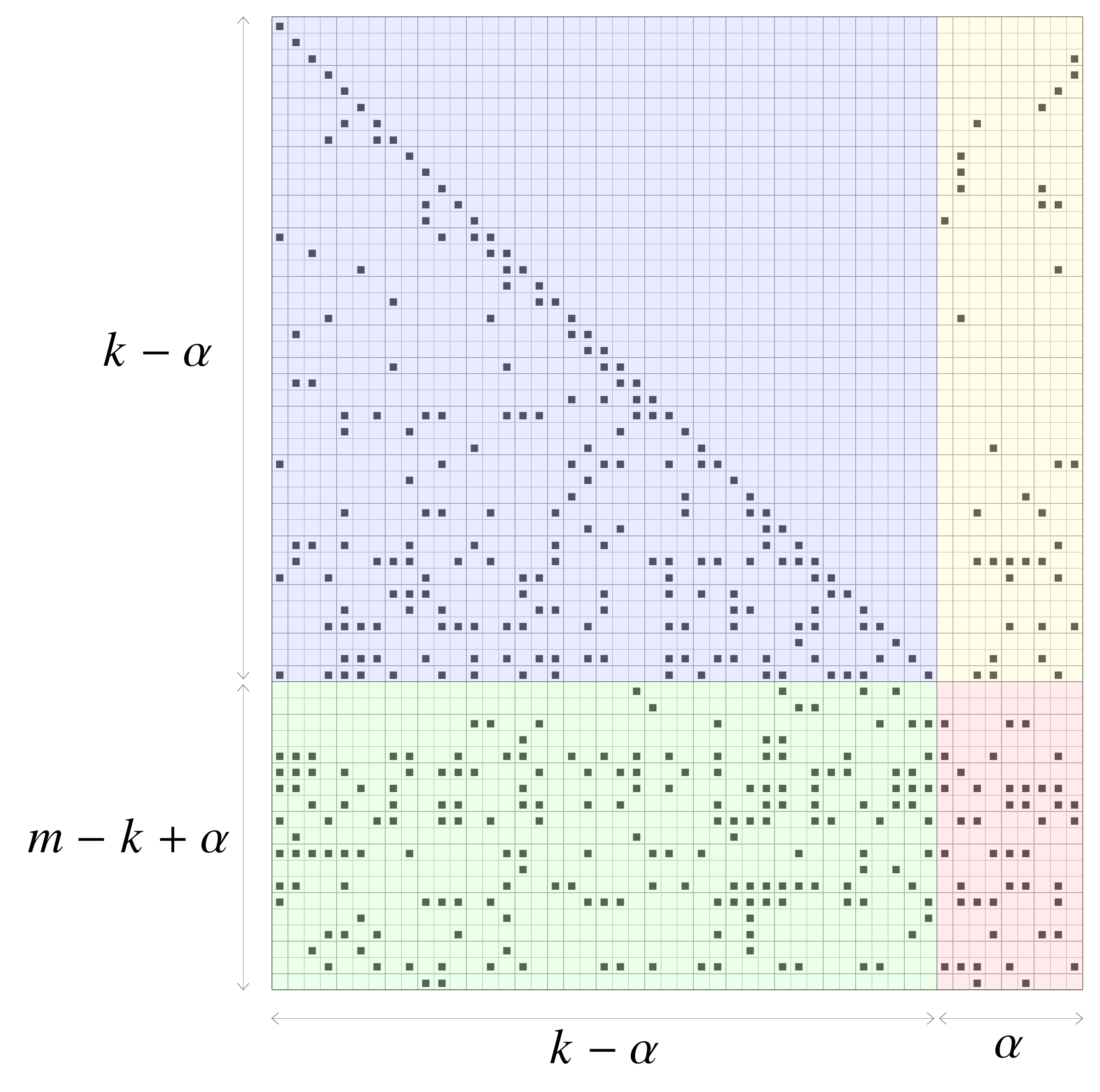}}\hspace{0.3cm}
	\subfloat[Zero matrix procedure.\label{fig:inactivation_2}]{\centering\includegraphics[height=0.55\columnwidth]{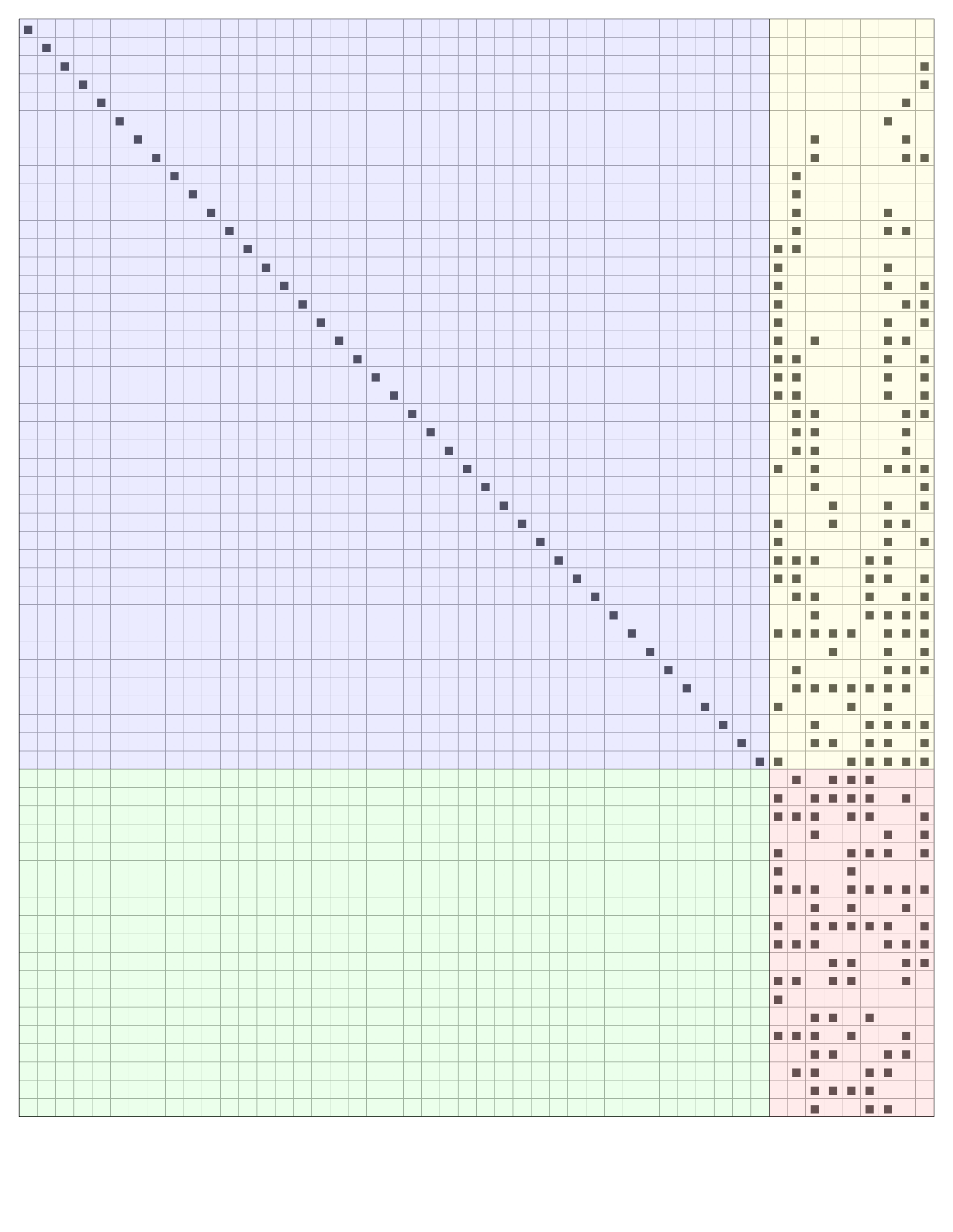}}\hspace{0.3cm}
	\subfloat[\acl{GE}.\label{fig:inactivation_3}]{\centering\includegraphics[height=0.55\columnwidth]{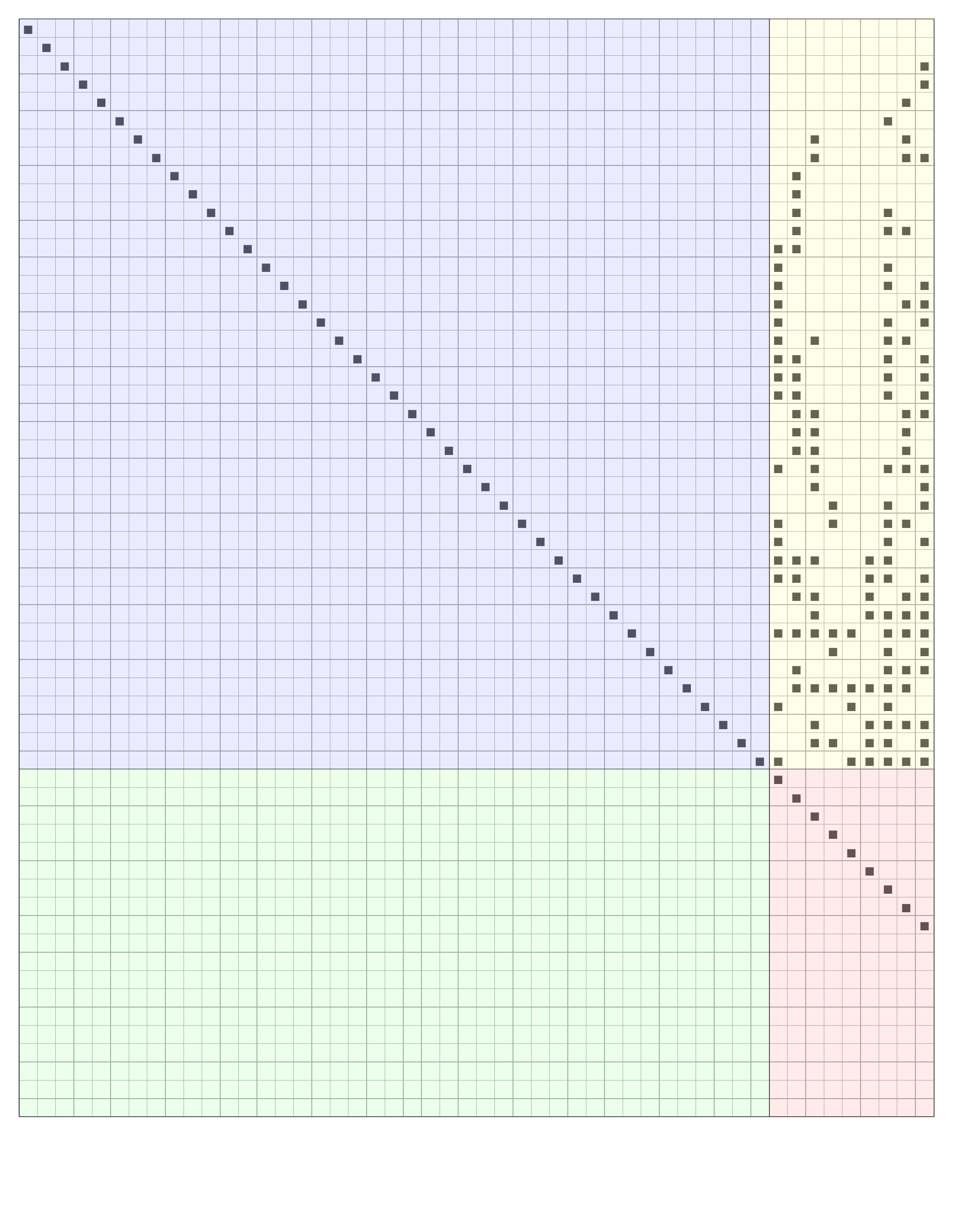}}\hspace{0.3cm}
	\subfloat[Back-substitution.\label{fig:inactivation_4}]{\centering\includegraphics[height=0.55\columnwidth]{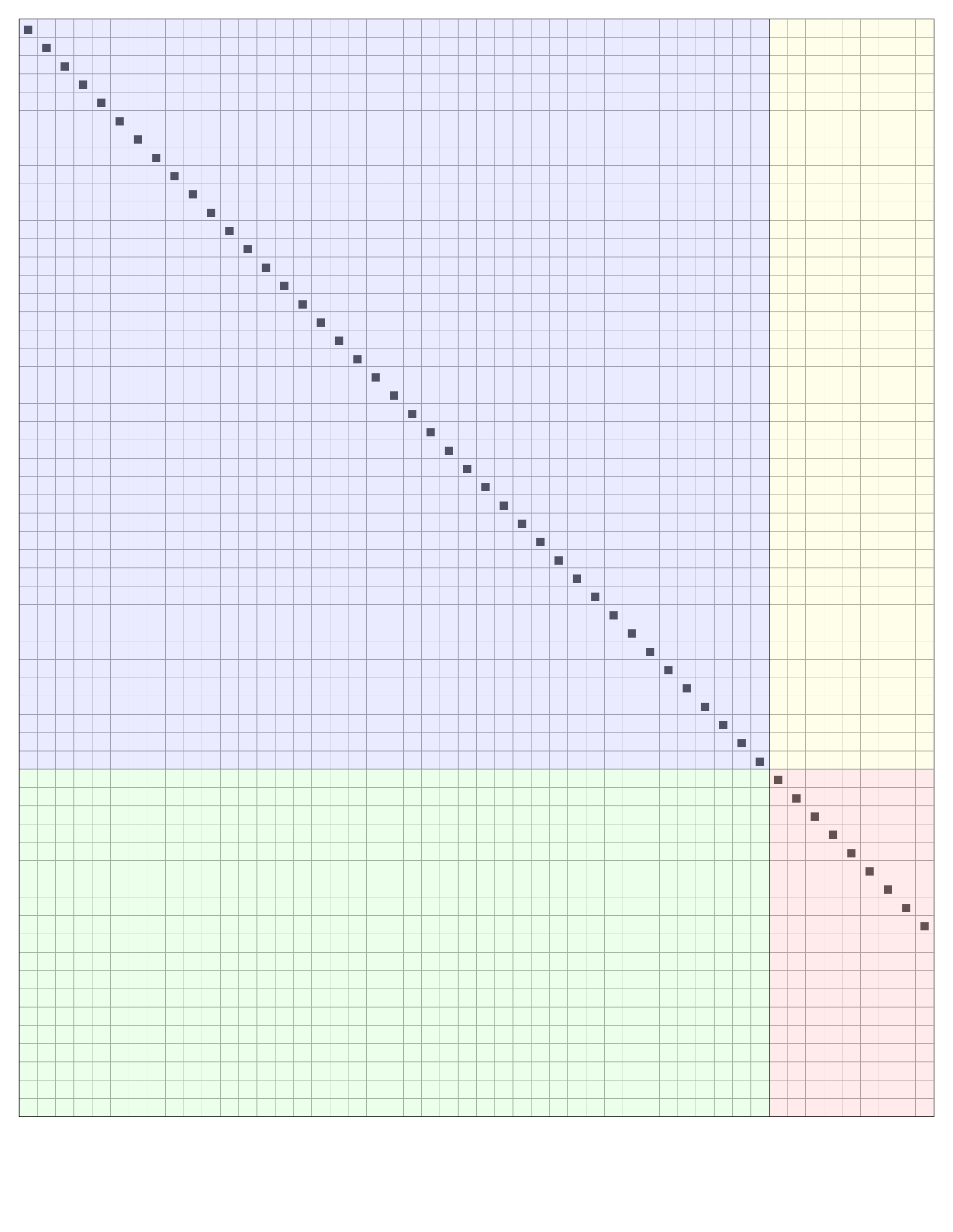}}	
	\caption{Structure of $\Grx^\t$ as inactivation decoding proceeds.}\label{fig:inactivation}
\end{figure*}

Inactivation decoding is an efficient \ac{ML} erasure decoding algorithm, which can be used to solve the system of equations in  \eqref{eq:ml_eq_sys}.
We will illustrate inactivation decoding  by means of an example and with the aid of Figures~\ref{fig:inactivation_0} and \ref{fig:inactivation}. In the example, we fix $k=50$ and $m=60$. The structure of $\Grx^\t$ is given in Figure~\ref{fig:inactivation_0} (in the figure, dark squares represent the non-zero elements of $\Grx^\t$). Inactivation decoding consists of $4$ steps.

\begin{step}[Triangulation]
	Initially, $\Grx^\t$ is put in an approximate lower triangular form by means of column and row permutations only.   Given that no operation is performed on the rows or columns of $\Grx^\t$, the density of $\Grx^\t$ is preserved. At the end of triangulation process, $\Grx^\t$ can be partitioned in four submatrices as
	\[
	\left[
	\begin{array}{c|c}
	\Amatrix & \Dmatrix \\ \hline
	\Bmatrix & \Cmatrix
	\end{array}
	\right]
	\]
	and as depicted in  Figure~\ref{fig:inactivation_1}. In the upper left part we have the $(k-\reducedsyst) \times (k-\reducedsyst)$ lower triangular matrix $\Amatrix$.
	The matrix under submatrix $\Amatrix$ is denoted by $\Bmatrix$ and it has dimension $ (m-k+\reducedsyst) \times (k-\reducedsyst)$.
	The  upper right part is given by the $(k-\reducedsyst) \times \reducedsyst$ submatrix $\Dmatrix$, while the lower right submatrix is denoted by $\Cmatrix$ and it has dimension $(m-k+\reducedsyst) \times \reducedsyst$. The $\reducedsyst$ rightmost columns of $\Grx^\t$ (corresponding to matrices $\Cmatrix$ and $\Dmatrix$) are usually referred to as inactive columns. Note that the manipulation of $\Grx^\t$ is typically performed through \emph{inactivation} or \emph{pivoting} algorithms (see e.g. \cite{miller04:bec,paolini12:TCOM}), which aim at reducing the number of inactive columns. In the rest of this work, we will assume that the use of the random inactivation algorithm \cite{paolini12:TCOM}, as it will be detailed later in the section.
\end{step}

\begin{step}[Zero matrix procedure]
	Matrix $\Amatrix$ is put in a diagonal form and matrix $\Bmatrix$ is zeroed out by means of row sums. As a result, on average the density of the matrices $\Cmatrix$ and $\Dmatrix$ increases (Figure~\ref{fig:inactivation_2}).
\end{step}

\begin{step}[Gaussian elimination]
	\ac{GE} is applied to solve the system of equations associated with $\Cmatrix$. Being $\Cmatrix$ in general dense, the complexity of this step is  cubic in the number of inactive columns $\reducedsyst$. As observed in \cite{miller04:bec}, this step drives, for large equation systems, the complexity of inactivation decoding.  After performing \ac{GE}, matrix $\Grx^\t$ has the structure shown in Figure~\ref{fig:inactivation_3}.	
\end{step}

\begin{step}[Back substitution]
	If Step 3 succeeds, after determining the value of the inactive variables (i.e., of the input symbols associated with the inactive columns), back-substitution is applied to compute the values of the remaining variables in $\mathbf{v}$. This is equivalent to setting to zero all elements of matrix $\Dmatrix$ in Figure~\ref{fig:inactivation_3}. At the end $\Grx^\t$ shows a diagonal structure as shown in Figure~\ref{fig:inactivation_4}, and all input symbols are recovered.
\end{step}

Recalling that a unique solution to the system of equations exists if and only if $\Grx$ is full rank, we have that decoding succeeds if and only if the rank of $\Cmatrix$ at Step 3 equals the number of inactive variables.

Given that the number of inactive variables is determined by the triangulation step, and that the \ac{GE}   on $\Cmatrix$ dominates the decoding complexity for large $k$, a first analysis of inactivation decoding can be addressed by focusing on the triangulation procedure.

\subsection{Bipartite Graph Representation of Inactivation Decoding}

We introduce next a bipartite graph representation of the \ac{LT} code from the receiver perspective. The graph comprises $m+k$ nodes, divided in two sets. The first set consists of $k$ \emph{input symbol nodes}, one per input symbol, while the second set consists of $m$ \emph{output symbol nodes}, one per received output symbol. We denote the set of input symbol nodes as $\inputset$, and the input symbol nodes in $\inputset$ as $v_1,v_2,\ldots,v_k$. Similarly, we denote the set of output symbol nodes as $\outputset$ with output symbol nodes denoted by $\rosymb_1,\rosymb_2,\ldots,\rosymb_m$. Here, we purposely referred to  input and output symbol nodes with the same notation used for their respective input and output symbols to emphasize the correspondence between the two sets of nodes and the sets of input and received output symbols.	For simplicity, in the following we will refer to input (output) symbol nodes and to input (output) symbols interchangeably.
An input symbol node $v_i$ is connected by an edge to an output symbol node $\rosymb_j$ if and only if the corresponding input symbol contributes to the generation of the corresponding output symbol, i.e., if and only if the $(i,j)$ element of $\Grx$ is equal to $1$.
We denote by $\deg(v)$ (or $\deg(\rosymb)$) the degree of a node $v$ (or $\rosymb$), i.e., the number of its adjacent edges.

Triangulation can be modelled as an iterative pruning of the bipartite graph of the \ac{LT} code. At each iteration, a reduced graph is obtained, which corresponds to a sub-graph of the original \ac{LT} code graph. The sub-graph involves only a subset of the original input symbols, and their neighboring output symbols. The input symbols in the reduced graph will be referred to as \emph{active} input symbols.  We shall use the term \emph{reduced} degree of a node to refer to the  degree of a node in the reduced graph. Obviously, the reduced degree of a node  is less than or equal to its original degree. We denote by $\rdeg(v)$ (or $\rdeg(\rosymb)$) the reduced degree of a node $v$ (or $\rosymb$). Let us now introduce some additional definitions that will be used to model the triangulation step.

\begin{mydef}[Ripple] We define the ripple as the set of output symbols of reduced degree 1 and we denote it by $\rippleset$.
\end{mydef}
\noindent The cardinality of the ripple is denoted by $\r$ and the corresponding random variable as $\Ripple$.

\begin{mydef}[Cloud] We define the cloud as the set of output symbols of reduced degree $d\geq 2$ and we denote it by $\cloudset$.
\end{mydef}
\noindent The cardinality of the cloud is denoted by $\c$ and the corresponding random variable as $\Cloud$.

\medskip

Initially, all input symbols are marked as active, i.e., the reduced sub-graph coincides with the original graph. At every step of the process, one active input symbol is  marked as either \emph{resolvable} or \emph{inactive} and leaves the graph. After $k$ steps no active symbols are present and triangulation ends. In order to keep track of the steps of the triangulation procedure,
the temporal dimension will be added through the subscript $u$.  This subscript $u$ corresponds to the number of active input symbols in the graph. Given the fact that the number of active input symbols decreases by $1$ at each step, triangulation will start with $u=k$ active input symbols and it will end after $k$ steps with $u=0$. Therefore, the subscript decreases as the triangulation procedure progresses.
Triangulation with random inactivations works as follows. Consider the transition from $u$  to $u-1$ active input symbols. Then,
\begin{itemize}
	\item If the ripple $ \ripple{u}$ is not empty $(\ru>0)$,
	the decoder selects an output symbol $\rosymb  \in \ripple{u}$ uniformly at random. The only  neighbor of $\rosymb$ is marked as resolvable and leaves the reduced graph, while all edges attached to it are removed.
	\item {If the ripple $\ripple{u}$ is empty $(\ru=0)$},
	one of the active input symbols $v$ is chosen uniformly at random\footnote{This is certainly neither the only possible inactivation strategy nor the one leading to the least number or inactivations. However, this strategy makes the analysis tractable. For an overview of the different inactivation strategies we refer the reader to \cite{paolini12:TCOM}.} and it is marked as inactive, leaving the reduced graph.  All edges attached to $v$ are removed.
\end{itemize}
{The input symbols marked as inactive are recovered using Gaussian elimination (step 3). After recovering the inactive input symbols, the remaining input symbols, those marked as resolvable, can be recovered by iterative decoding (back substitution in step 4).}

\begin{example}
	We provide an example for an \ac{LT} code with $k=4$ input symbols and $m=4$  output symbols (Figure~\ref{fig:example}).
	\begin{figure*}
		\centering        \subfloat[$u=4$.\label{fig:example_4}]{\centering\includegraphics[width=0.51\columnwidth]{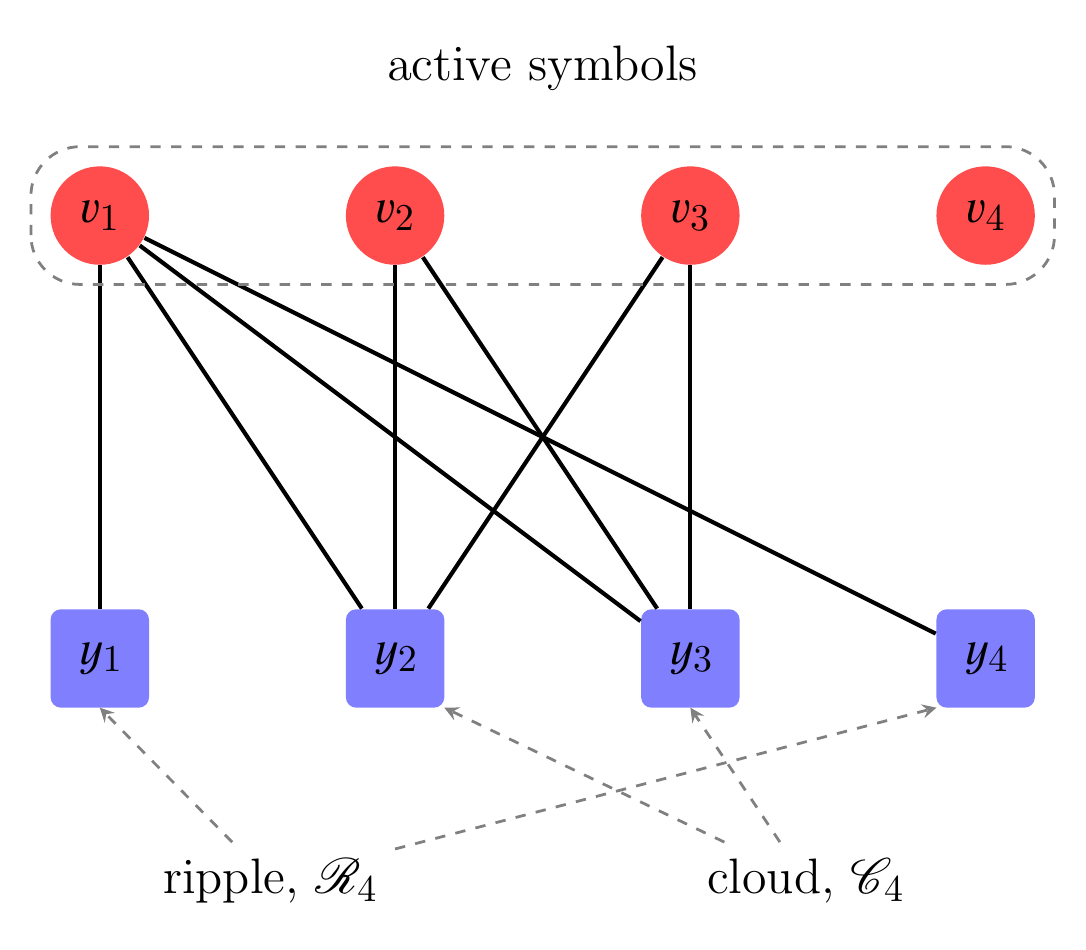}}\hspace{0.9cm}
		\subfloat[$u=3$.\label{fig:example_3}]{\centering\includegraphics[width=0.51\columnwidth]{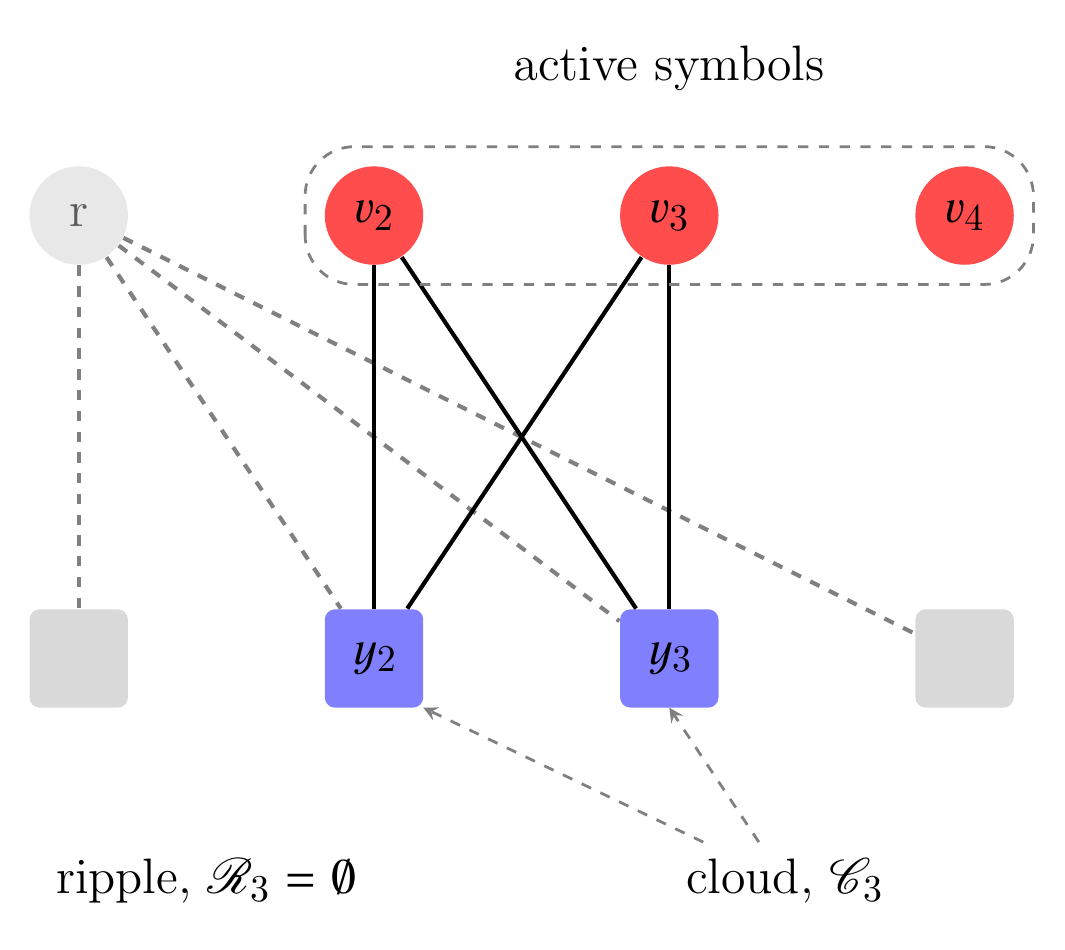}}\hspace{0.9cm}
		\subfloat[$u=2$.\label{fig:example_2}]{\centering\includegraphics[width=0.51\columnwidth]{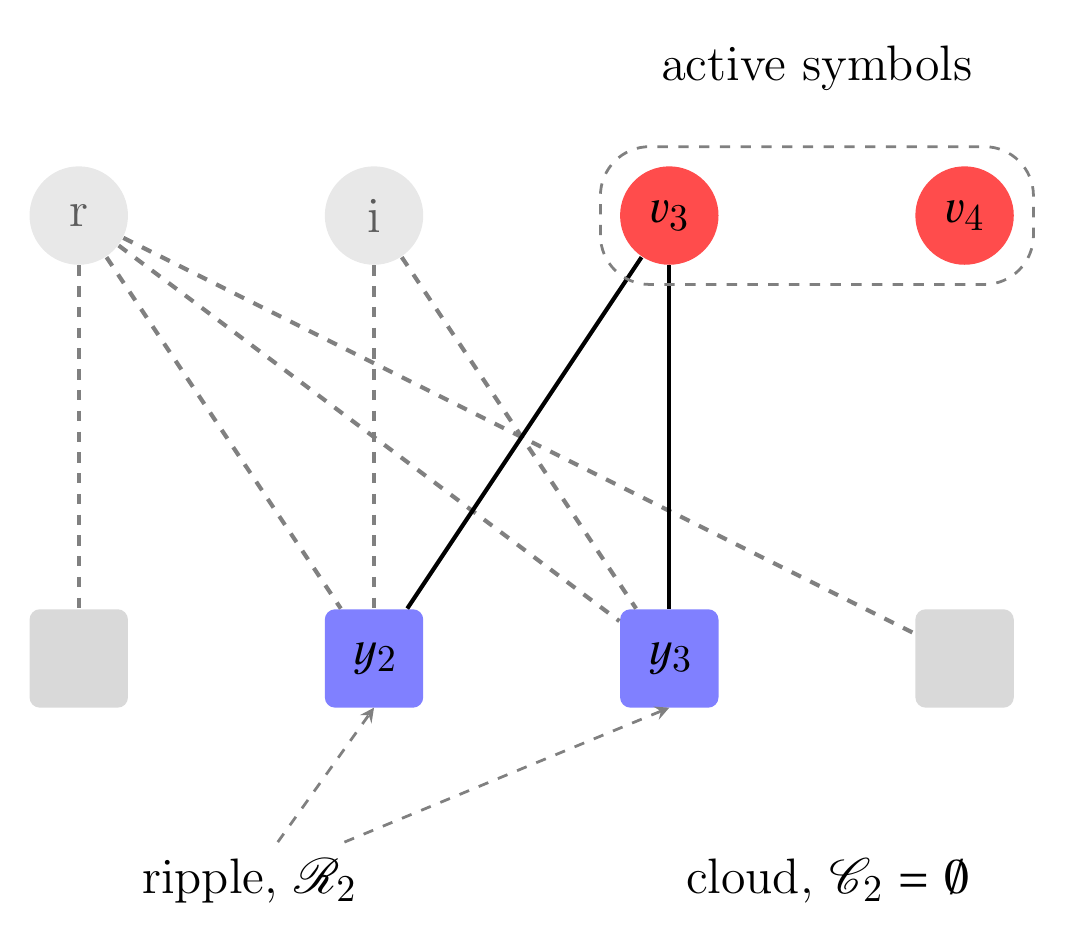}}\\[0.5cm]
		\subfloat[$u=1$.\label{fig:example_1}]{\centering\includegraphics[width=0.55\columnwidth]{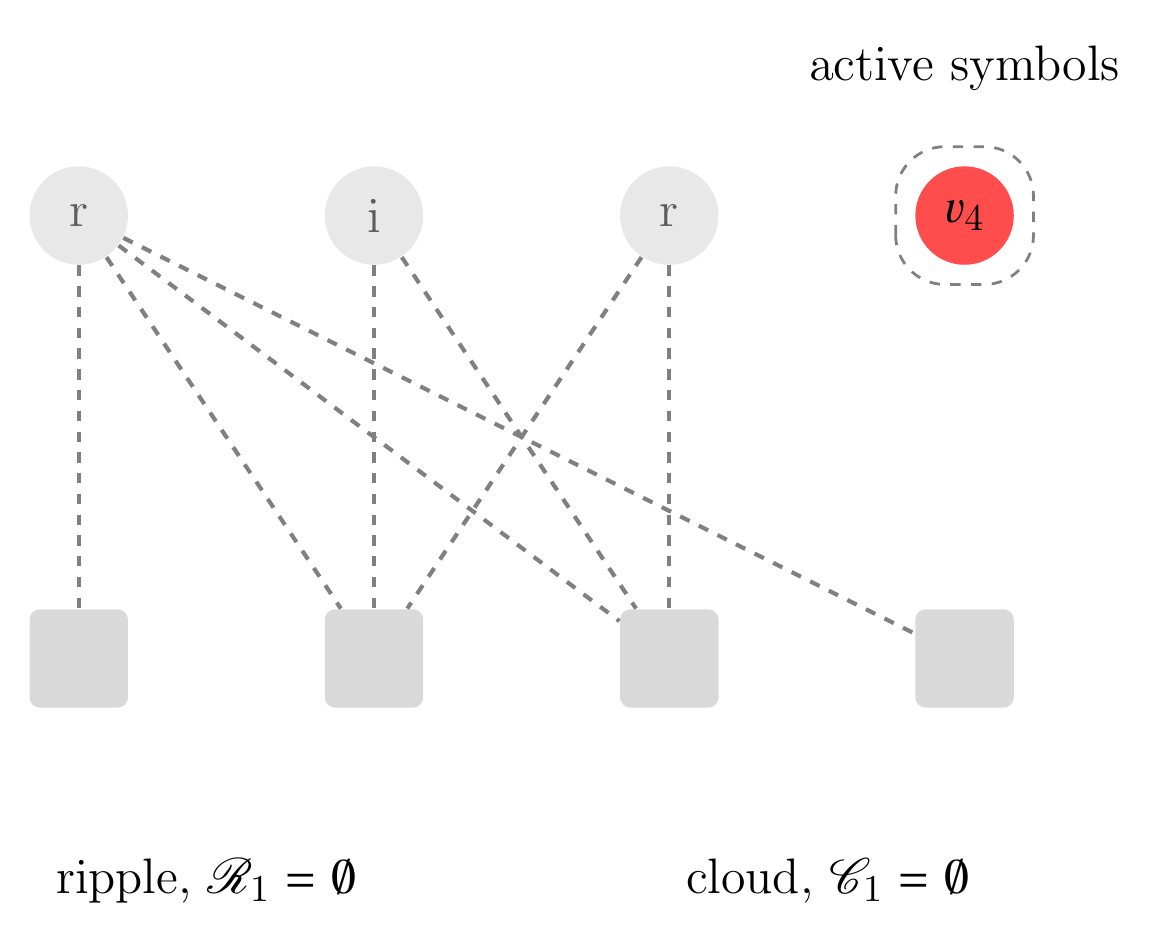}}\hspace{0.9cm}
		\subfloat[$u=0$.\label{fig:example_0}]{\centering\includegraphics[width=0.50\columnwidth]{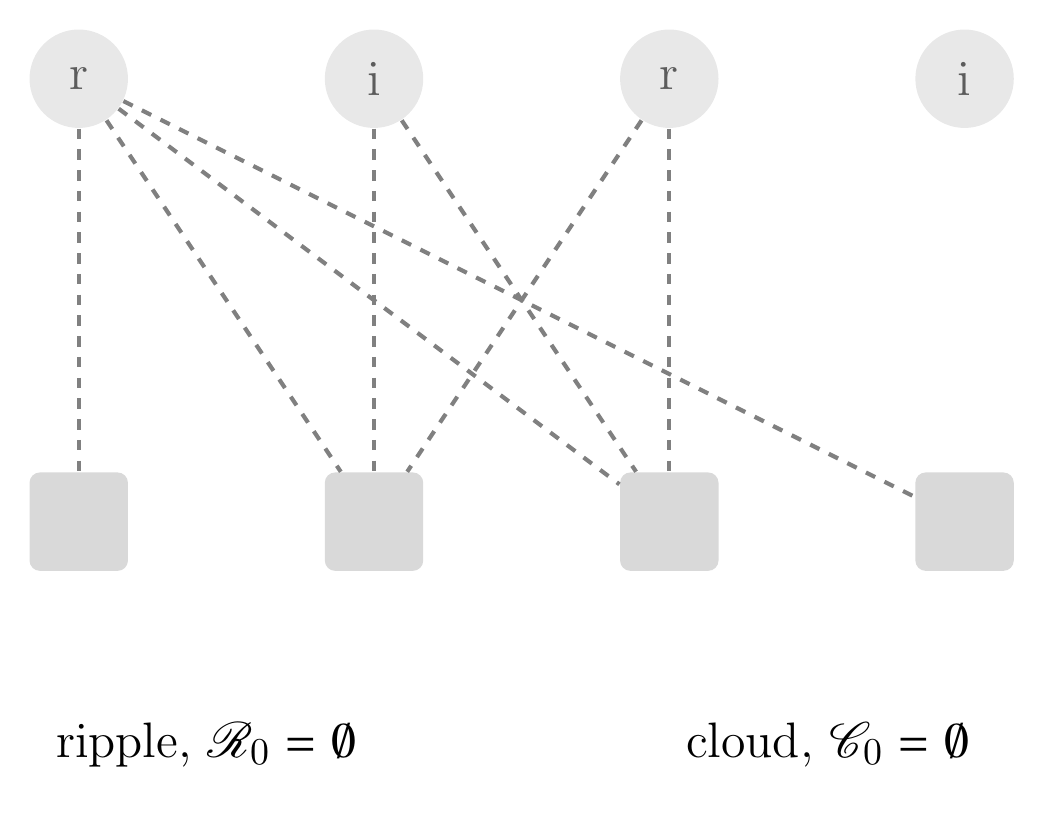}}\\[0.5cm]
		\caption{Triangulation procedure example}\label{fig:example}
	\end{figure*}
	\begin{enumerate}
		\item[i.] {Transition from $u=4$ to $u=3$. Initially, there are two output symbols in the ripple ($\r_{4}=2$) (Figure~\ref{fig:example_4}). Thus, one of the input symbols in the ripple is randomly selected and marked as resolvable. In this case symbol $v_1$ is selected and all its adjacent edges are removed. The graph obtained after the transition is shown in  Figure~\ref{fig:example_3}. Observe that the nodes $\rosymb_1$ and $\rosymb_4$ have left the graph since their reduced degree is now zero.}
		\item[ii.] {Transition from $u=3$ to $u=2$. As shown in Figure~\ref{fig:example_3} the ripple is now empty ($\r_{3}=0$). Thus, an inactivation has to take place. Node $v_2$ is chosen (according to a random selection) and is marked as inactive. All edges attached to $v_2$ are removed from the graph. As a consequence, the nodes $\rosymb_2$ and $\rosymb_3$ that were in the cloud $\cloudset_3$  enter the ripple $\rippleset_2$ (i.e., their reduced degree is now $1$), as shown in  Figure~\ref{fig:example_2}.}
		\item[iii.]{Transition from $u=2$ to $u=1$. Now the ripple is not empty ($\r_{2}=2$). The input symbol $v_3$ is selected (again, according to a random choice) from the ripple and is marked as resolvable. All its adjacent edges are removed. The nodes $\rosymb_2$ and $\rosymb_3$ leave the graph because their reduced degree becomes zero (see Figure~\ref{fig:example_1}).}
		\item[iv.] {Transition from $u=1$ to $u=0$. The ripple is now empty (Figure~\ref{fig:example_1}). Hence, an inactivation takes place: node $v_4$ is marked as inactive and the triangulation procedure ends.}
	\end{enumerate}
	Note that in this example input symbol $v_4$ has no neighbors, i.e., the matrix $\Grx$ is not full rank and decoding fails.
\end{example}

\section{Analysis under Random Inactivation} \label{chap:inact_analysis}

In this section we analyze the triangulation procedure with the objective of determining the average number of inactivations, i.e., the expected number of inactive input symbols at the end of the triangulation process. We will then extend the analysis to obtain the probability distribution of the number of inactivations.

\subsection{Average Number of Inactivations} \label{chap:inact_first_order}

Following \cite{Karp2004,shokrollahi2009theoryraptor,Maatouk:2012}, we model the decoder as a finite state machine with state at time $u$ given by the cloud and the ripple sizes at time $u$, i.e.
\[
\S{u}:=(\Cu, \Ru ).
\]
We aim at deriving a recursion for the decoder state, that is, to obtain ${\Pr \{ \S{u-1}=(\c_{u-1}, \r_{u-1}) \}}$  as a function of $\Pr \{ \S{u}=(\cu, \ru )\}$. We do so by analyzing how the ripple and cloud change in the transition from $u$ to $u-1$.
In the transition exactly one active input symbol is marked as either resolvable or inactive and all its adjacent edges are removed. Whenever edges are removed from the graph, the reduced degree of one or more output symbols decreases. Consequently, some of symbols in the cloud may enter the ripple and some of the symbols in the ripple may become of reduced degree zero and leave the graph.

\medskip

We focus first on the symbols that leave the cloud and enter the ripple during the transition at step $u$,  conditioned on ${\S{u}=(\c_u, \r_u)}$.  Since for an \ac{LT} code the neighbors of the output symbols are selected independently and uniformly at random, in the transition each output symbol may leave the cloud and enter the ripple independently from the other output symbols. Hence,
the number of symbols leaving $\cloud{u}$ and entering $\ripple{u-1}$ is binomially distributed with parameters $\c_u$ and $\pru$ with
\begin{align}
	\pru :&= \Pr \{ \rosymbrv \in \ripple{u-1} | \rosymbrv \in \cloud{u} \}\\ &= \frac { \Pr \{ \rosymbrv \in \ripple{u-1}\, , \, \rosymbrv \in \cloud{u} \} }  { \Pr \{ \rosymbrv \in \cloud{u} \}}
	\label{eq:pu_prob}
\end{align}
where random variable $\rosymbrv$ represents a randomly chosen output symbol.

We first consider the numerator of \eqref{eq:pu_prob}. Conditioning on the original degree of $\rosymbrv$, we have the following proposition.
\begin{prop}\label{prop:1}
	The probability that an output symbol $\rosymbrv$ belongs  to the cloud at step $u$  and  enters the ripple at step $u-1$,  condition to its original degree being $d$, is
	\begin{align}
		\Pr &\{ \rosymbrv \in \ripple{u-1}\, , \,  \rosymbrv \in \cloud{u} | \deg(\rosymbrv)= d \} =\\   &\begin{dcases}
			(u-1)  \binom{k-u}{d-2}\binom{k}{d}^{-1} & \text{if }  2 \leq d \leq k-u{+}2,\\
			0 & \text{otherwise.}
		\end{dcases}
		\label{eq:z_and_l_d}
	\end{align}
\end{prop}
\begin{IEEEproof}
	For an output symbol $\rosymbrv$ of degree $d$ to belong  to the cloud at step $u$  and  to the ripple at step $u-1$ we need that the output symbol has \emph{reduced} degree $2$ \emph{before} the transition and \emph{reduced} degree $1$ \emph{after} the transition. For this to happen {two events must take place:
		\begin{itemize}
			\item $\eventa$ one of the $d$ edges of output symbol $\rosymbrv$ is connected to the symbol being marked as inactive or resolvable at the transition,
			\item $\eventb$  another edge is connected to one of the $u-1$ active symbols after the transition
			and at the same time the remaining $d-2$ edges connected to the $k-u$ not active input symbols (inactive or resolvable).
		\end{itemize}}
		{The joint probability of $\eventa$ and $\eventb$ can be derived as
			\[
			\Pr\{\eventa , \eventb\} = \Pr\{\eventa \} \Pr\{\eventb | \eventa\}
			\]
			Let us focus first on $\eventa$. We consider an output symbols of degree $d$, thus this is simply probability that one the $d$ edges of $\rosymbrv$ is connected to the symbol being marked as inactive or resolvable at the transition,  $\Pr\{\eventa\}=d/k$.
			\newline
			Let us now focus on  $\Pr\{\eventb | \eventa\}$. Since we condition on  $\eventa$ we have that one of the $d$ edges of $\rosymbrv$ is already assigned to one input symbol. We now need to consider the remaining $d-1$  edges and $k-1$ input symbols.
			The probability that one of the $d-1$ edges is connected to the set of $u-1$ active symbols is $(d-1)(u-1)/(k-1)$.
			At the same time we must have exactly $d-2$ edges going to the $k-u$ input symbols that were not active before the transition. Hence, we have
			\[
			\Pr\{\eventb | \eventa\} = (d-1)\frac{u-1}{k-1} \binom{k-u}{d-2} {\binom{k-2}{d-2}}^{-1}
			\]
			By multiplying the two probabilities, applying few manipulations, and making explicit the conditioning on $\deg(\rosymbrv)= d$, we get}
		\[
		\Pr \{ \rosymbrv \in \ripple{u-1}\, , \,  \rosymbrv \in \cloud{u} | \deg(\rosymbrv)= d \} = (u-1)  \binom{k-u}{d-2}\binom{k}{d}^{-1}.
		\]
		We shall anyhow observe that if $\deg(\rosymbrv) < 2$, the output symbol cannot belong to the cloud. Moreover, one needs to impose the condition $d-2 \leq k-u$, since output symbols choose their neighbors without replacement (an output symbol cannot have more than one edge going to an input symbol). Hence, for $d<2$ and $d>k-u+2$, the probability $\Pr \{ \rosymbrv \in \ripple{u-1}\, , \,  \rosymbrv \in \cloud{u} | \deg(\rosymbrv)= d \}$ is zero.
	\end{IEEEproof}
	
	\medskip
	
	By
	removing the conditioning on the degree of $\rosymbrv$ in \eqref{eq:z_and_l_d},  we have
	\begin{align}
		\Pr \{ \rosymbrv  \in \ripple{u-1}\, , \,  \rosymbrv  \in \cloud{u} \} =
		(u-1)\mathlarger {\sum}_{d=2}^{k-u+2}   \Omega_d  {\binom{k-u}{d-2}}{\binom{k}{d}^{-1}}.
		\label{eq:z_and_l}
	\end{align}
	
	\medskip
	
	Let us now focus on the denominator of \eqref{eq:pu_prob} which gives the probability that a randomly chosen output symbol $\rosymbrv$ is in the cloud when $u$ input symbols are still active. This probability is provided by the following proposition.
	\begin{prop}\label{prop:2}
		The probability that the randomly chosen output symbol $\rosymbrv$ is in the cloud when $u$ input symbols are still active is
		\begin{align}
			\Pr  \{ \rosymbrv \in \cloud{u}\}=  1 &- u\mathlarger{\sum}\limits_{d=1}^{{k-u+1}}   \Omega_d  \binom{k-u}{d-1}\binom{k}{d}^{-1}+\\
			&-  \mathlarger{\sum}\limits_{d=1}^{{k-u}}   \Omega_d \binom{k-u}{d}\binom{k}{d}^{-1}.
			\label{eq:z}
		\end{align}
	\end{prop}
	\begin{IEEEproof}
		The probability of $\rosymbrv$ not being in the cloud is given by the probability of $\rosymbrv$ having reduced degree $0$ or being in the ripple. Given that the two events are mutually exclusive, we can compute such probability as the sum of the probabilities of the two events,
		\begin{align}
			\Pr \{ \rosymbrv \in \cloud{u} \}
			&=  1 -  \Pr \{ \rosymbrv \in \ripple {u} \cup  \rdeg_u (\rosymbrv) = 0 \} \\
			&=1 -  \Pr \{ \rosymbrv \in \ripple{u} \} - \Pr \{ \rdeg_u (\rosymbrv) = 0 \}\label{eq:PrYinCloudu}
		\end{align}
		where $\rdeg_u (\rosymbrv)$ denotes the reduced degree of output symbol $\rosymbrv$ when $u$ input symbols are still active.
		We first focus on the probability of $\rosymbrv$ being in the ripple. Let us assume $\rosymbrv$ has original degree $d$. The probability that $\rosymbrv$ has reduced degree $1$ equals the probability of $\rosymbrv$ having exactly one neighbor among the $u$ active input symbols and the remaining $d-1$ neighbors among the $k-u$ non-active (i.e., solved or inactive) ones. Thus, we have
		\begin{equation}
			\Pr \{ \rosymbrv \in \ripple {u}| \deg(\rosymbrv)=d \}
			= d \frac{u}{k} \frac{\binom{k-u}{d-1}}{ \binom{k-1}{d-1}}=u\binom{k-u}{d-1}\binom{k}{d}^{-1}. ~~
			\label{eq:prop_in_1}
		\end{equation}
		The probability of $\rosymbrv$ having reduced degree $0$ is the probability that all $d$ neighbors of $\rosymbrv$ are in the $k-u$ non-active symbols. The total number of different edge assignments is $\binom{k}{d}$, and the total number of edge assignments in which all $d$ neighbors of $\rosymbrv$ are in the $k-u$ non-active symbols is $\binom{k-u}{d}$. Thus we have
		\begin{equation}
			\Pr \{ \rdeg_u (\rosymbrv) = 0| \deg(\rosymbrv)=d \} = \binom{k-u}{d}\binom{k}{d}^{-1}.
			\label{eq:prop_in_2}
		\end{equation}
		
		By removing the conditioning on $d$ in \eqref{eq:prop_in_1} and \eqref{eq:prop_in_2} and by replacing the corresponding results in \eqref{eq:PrYinCloudu} we obtain \eqref{eq:z}.
	\end{IEEEproof}
	
	\medskip
	
	The expression of $\pru$ is finally given in the following proposition.
	
	\begin{prop}
		The probability $\pru$ that a randomly chosen output symbol leaves the cloud $\cloud{u}$ enters the ripple $\ripple{u-1}$ after the transition from $u$ to $u-1$ active input symbols is
		\begin{align}
			\pru =  \frac{ (u-1)\mathlarger{\sum}\limits_{d=2}^{k-u+2}  \Omega_d    \binom{k-u}{d-2}\binom{k}{d}^{-1}}
			{  1 - u\mathlarger{\sum}\limits_{d=1}^{{k-u+1}} \Omega_d  \binom{k-u}{d-1}\binom{k}{d}^{-1}
				-  \mathlarger{\sum}\limits_{d=1}^{{k-u}}\Omega_d \binom{k-u}{d}\binom{k}{d}^{-1}    }.
		\end{align}
	\end{prop}
	\begin{IEEEproof}
		The proof follows directly from \eqref{eq:pu_prob} and Propositions~\ref{prop:1} and \ref{prop:2}.
	\end{IEEEproof}
	
	\medskip
	
	Let us now focus on the number  $\erv_u$ of symbols leaving the ripple during the transition from $u$ to $u-1$ active symbols. We denote by $\Erv_u$ the random variable associated with $\erv_u$. Two cases shall be considered.
	In a first case, no inactivation takes place because the ripple is not empty. Thus, an output symbol $\rosymbrv$ is chosen at random from the ripple and its only neighbor $v$ is marked as resolvable and removed from the graph. By removing $v$ from the graph, other output symbols that are connected to $v$ and that are in the ripple leave the ripple. Thus, for $\ru>0$ we have
	\begin{align}
		\Pr\{\Erv_u=\erv_u & |\Ru=\ru\} = \\
		&\binom{\ru-1}{\erv_u-1} \left(\frac{1}{u}\right)^{\erv_u-1} \left( 1- \frac{1}{u} \right)^{\ru-\erv_u}
		\label{eq:erv_non_empty}
	\end{align}
	with $1\leq \erv_u \leq \ru$.
	In the second case, the ripple is empty ($\ru=0$) and an inactivation takes place. Since no output symbols can leave the ripple, we have
	\begin{align}
		\Pr\{\Erv_u=\erv_u|\Ru=0\} = \begin{cases} 1  & \text{if } \erv_u = 0 \\ 0  & \text{if } \erv_u > 0. \end{cases}
		\label{eq:erv_empty}
	\end{align}
	We are now in the position to derive the transition probability
	$\Pr\{\S{u-1}=(\c_{u-1},\r_{u-1})|\S{u}=(\c_{u},\r_{u})\}$.
	Let us introduce the cloud drift $\b_u$ to denote the variation of number of cloud elements in the transition from $u$ to $u-1$ active symbols, i.e.,
	\[
	\b_u:=\cu-\c_{u-1}.
	\]
	We can now express the cloud and ripple cardinality after the transition respectively as
	\begin{align*}
	\c_{u-1} &= \cu - \b_u\\
	\r_{u-1} &= \ru-\erv_u+\bu.
	\end{align*}
	{Since output symbols are generated independently,} the random variable associated to $\b_u$ is binomially distributed with parameters $\cu$ and $\pru$, and making use of \eqref{eq:erv_non_empty}, we obtain the following recursion
	\begin{align}
		\Pr\{\S{u-1}&=(\cu-\bu,\ru-\erv_u+\bu) | \S{u}=(\cu,\ru)\}  \\[2mm]
		&= \binom{\cu}{\bu} \pru^{\bu} (1-\pru)^{\cu-\bu}\\
		&\times \binom{\ru-1}{\erv_u-1}
		\left(\frac{1}{u}\right)^{\erv_u-1} \left( 1- \frac{1}{u} \right)^{\ru-\erv_u}
		\label{eq:prob_transition}
	\end{align}
	which is valid for $\ru>0$, while for $\ru=0$, according to \eqref{eq:erv_empty}, we have
	\begin{align}
		\Pr\{\S{u-1}&=(\cu-\bu,\bu) | \S{u}=(\cu,0)\} \\
		&= \binom{\cu}{\bu} {\pru}^{\bu} (1-\pru)^{\cu-\bu}.
		\label{eq:prob_transition_r_0}
	\end{align}
	Note that the probability of ${\S{u-1} =(\c_{u-1},\r_{u-1})}$ can be computed recursively via \eqref{eq:prob_transition}, \eqref{eq:prob_transition_r_0} by initializing  the decoder state as
	\begin{align}
		\Pr\{\S{k} =(\c_{k},\r_{k}) \} = \binom{m}{\r_{k}} \Omega_1^{\r_{k}} \left( 1-\Omega_1\right)^{\c_{k}}
	\end{align}
	for all non-negative $\c_{k},\r_{k}$ such that $\c_{k}+\r_{k} = m$, where $m$ is the number of  output symbols.
	
	The following proposition establishes how the number of inactivations can be determined using the finite state machine.
	\begin{theorem} \label{theorem:1}
		Let $\Y$ denote the random variable associated to the number of inactivations  at the end of the triangulation process. The expected value of $\Y$  is given by
		\begin{align}
			\Exp\left[\Y\right]= \sum_{u=1}^{k} \sum_{\cu}  \Pr\{\S{u} =(\cu,0) \}. \label{eq:avg_inact}
		\end{align}
	\end{theorem}
	\begin{IEEEproof}
		Let us denote by $\mathbf{\f}=(\f_1, \f_2, \cdots, \f_k)$ the binary vector associated to the inactivations performed during the triangulation process, and let $\mathbf{\F}=(\F_1, \F_2, \cdots, \F_k)$ denote the associated random vector. In particular, the $u$-th element of $\mathbf{\f}$, $\f_u$ is set to $1$ if an inactivation is performed when $u$ input symbols are active, and it is set to $0$ if no inactivation is performed. Thus, for a given instance of inactivation decoding, the total number of inactivations corresponds simply to the Hamming weight of $\mathbf{\f}$, which we denote as $w_H( \mathbf{\f} )$. The expected number of inactivations can be obtained as
		\begin{align}
			\Exp \left[ \Y \right] &= \sum_{ \mathbf{\f} }  w_H(\mathbf{\f}) \Pr\left\{ \mathbf{\F}=\mathbf{\f} \right\} \\
			&= \sum_{ \mathbf{\f} } \left( \sum_{u} \f_u \right) \Pr\left\{ \mathbf{\F}=\mathbf{\f} \right\}  = \sum_{u} \sum_{ \mathbf{\f} } \f_u \Pr\left\{ \mathbf{\F}=\mathbf{\f} \right\}
		\end{align}
		where the summation is taken over all possible vectors $\mathbf{\f}$. We shall now define  ${\mathbf{\f}_{\setminus u} = (\f_1, \cdots, \f_{u-1}, \f_{u+1}, \f_k)}$, i.e., $\mathbf{\f}_{\setminus u}$ denotes a vector containing all but the $u$-th element of $\mathbf{\f}$. We have
		\begin{align}
			\Exp \left[ \Y \right] &= \sum_{u} \sum_{ \mathbf{\f}_{ \setminus u }} \sum_{ \f_u } \f_u  \Pr\left\{  \mathbf{\F}=\mathbf{\f} \right\} \\
			&= \sum_{u} \sum_{ \f_u } \f_u  \sum_{ \mathbf{\f}_{ \setminus u }}   \Pr\left\{  \mathbf{\F}=\mathbf{\f} \right\} \\
			&= \sum_{u} \sum_{ \f_u } \f_u   \Pr\left\{ \F_u = \f_u \right\} \\
			&= \sum_{u} \Pr\left\{ \F_u=1 \right\}.
		\end{align}
		If we now observe that by definition
		\[
		\Pr\left\{ \F_u=1 \right\} = \sum_{\cu}  \Pr\{\S{u} =(\cu,0) \}
		\]
		we obtain \eqref{eq:avg_inact}, and the proof is complete.
	\end{IEEEproof}

\begin{figure}
		\begin{center}
			\includegraphics[width=\figw]{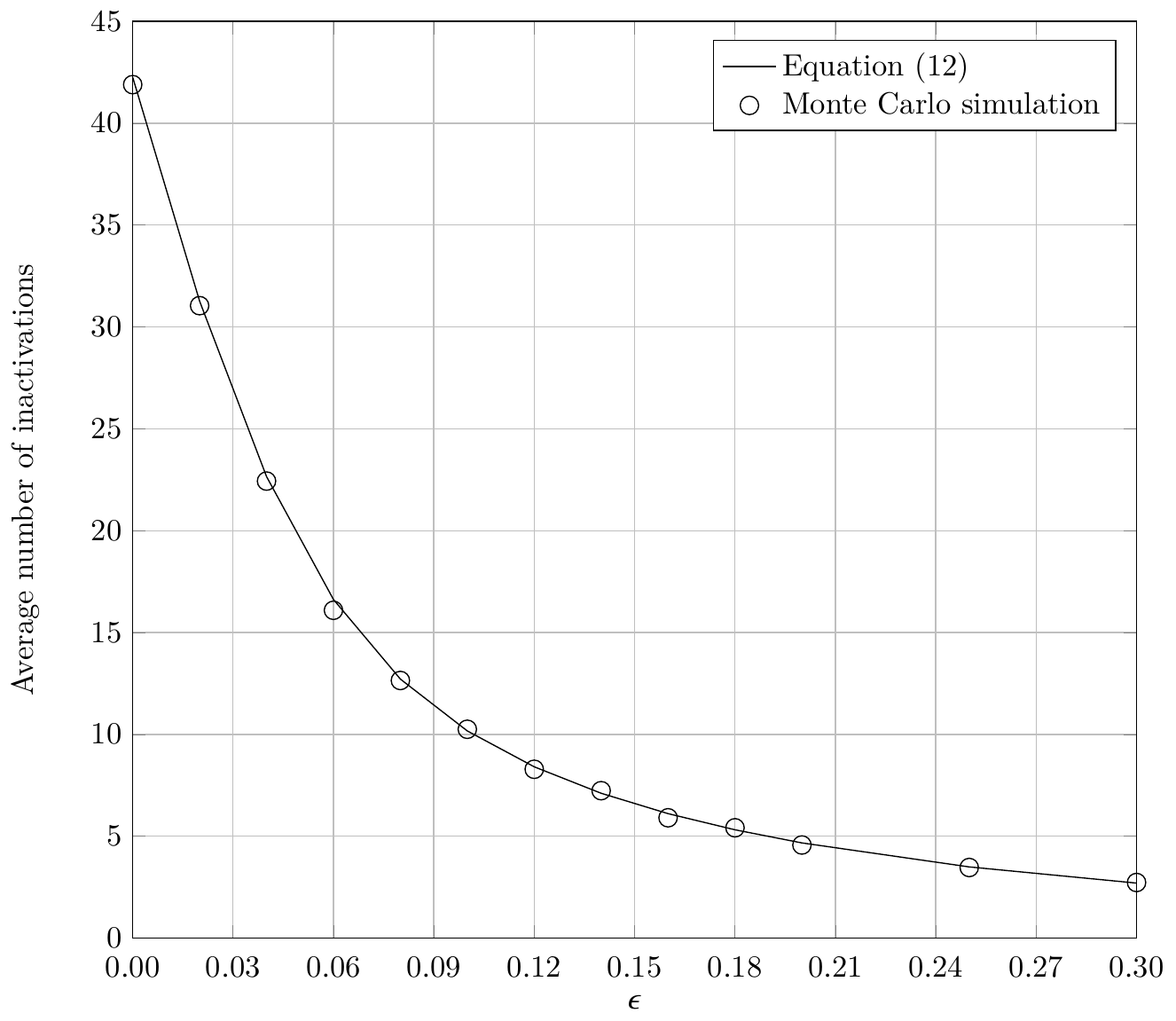}
			\centering \caption{Average number of inactivations vs.  relative overhead $\reloverhead$ for an \ac{LT} code with $k=1000$ and  with  degree distribution $\Omegaone(\x)$.}
			\label{fig:mbms_k_1000}
		\end{center}
	\end{figure}

	Figure ~\ref{fig:mbms_k_1000} shows the expected number of inactivations for a $k=1000$ \ac{LT} code with the output degree distribution
	\begin{align}
		\Omegaone(\x):& = 0.0098\x + 0.4590\x^2+ 0.2110\x^3+0.1134\x^4+ \\
		&+0.1113\x^{10} + 0.0799\x^{11} + 0.0156\x^{40}
		\label{eq:deg_dist_mbms}
	\end{align}
	which is the one adopted by standardized Raptor codes \cite{MBMS12:raptor,luby2007rfc}.
	The figure shows the number of inactivations according to \eqref{eq:avg_inact} and results obtained through Monte Carlo simulations. In particular, in order to obtain the simulation results for each value of relative overhead 1000 decoding attempts were carried out. A tight match between the analysis and the simulation results can be observed.
	
	{The results in this section are strongly based on \cite{shokrollahi2009theoryraptor}, where LT codes under inactivation decoding were analyzed. The difference is that, when considering the triangulation step of inactivation decoding, the decoding process does not stop when the decoder is at state $\S{u} =(\cu,0)$. Instead, decoding can be resumed after performing an inactivation.
	}
	
	\subsection{Distribution of the Number of Inactivations}\label{chap:inact_distribution}
	
\begin{figure}
		\begin{center}
			\includegraphics[width=\figw]{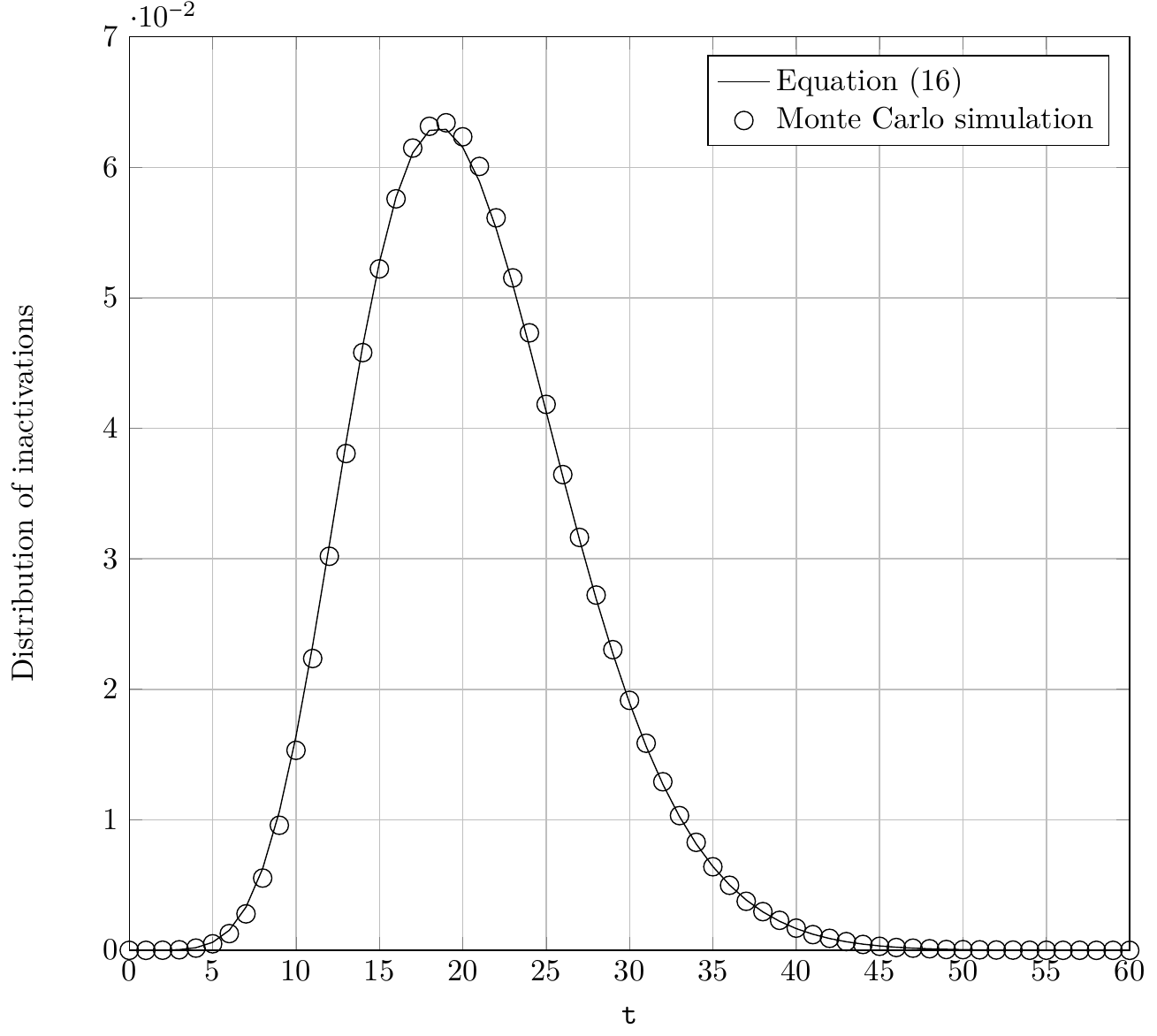}
			\centering \caption{Distribution of the number of inactivations for an \ac{LT} code with $k=300$, {relative overhead $\reloverhead = 0.02$} and  degree distribution $\Omegaone(\x)$ given in \eqref{eq:deg_dist_mbms}. }
			\label{fig:mbms_dist_inact}
		\end{center}
	\end{figure}

	The analysis presented in Section~\ref{chap:inact_first_order}  provides the expected number of inactivations at the end of the triangulation process under random inactivation decoding. In this section we extend the analysis to obtain the probability distribution of the number of inactivations.
	To do so, we extend the finite state machine by including the number  of inactive input symbols in the state definition, i.e.,
	\[
	\S{u}=(\Cu, \Ru, \Nu )
	\]
	where $\Nu$ is the random variable associated to the number of inactivations at step $u$ (when $u$ input symbols are active). We  proceed by deriving a recursion to obtain ${\Pr\{ \S{u-1}=(\c_{u-1}, \r_{u-1}, \n_{u-1} )\}}$ as a function of ${\Pr\{ \S{u}=(\cu, \ru, \nu )\}}$.
	Two cases shall be considered. When the ripple is not empty ($\r_{u}>0$) no inactivation takes place, at the transition from $u$ to $u-1$ active symbols the number of inactivations is unchanged ($\n_{u-1} = \nu$). Hence, we have
	\begin{align}
		\Pr\{& \S{u-1}=(\c_{u}  -\b_{u}, \r_{u}-\erv_{u}+\b_{u}, \nu) | \S{u}=(\c_u,\r_u,\nu)\} \\[-2mm]
		& =\binom{\c_u}{\b_u} {\pru}^{b_u} (1-\pru)^{\c_u-\b_u} \\[2mm]
		&\times \binom{\r_u-1}{\erv_u-1} \left(\frac{1}{u}\right)^{\erv_u-1} \left( 1- \frac{1}{u} \right)^{\r_u-\erv_u}.
		\label{eq:prob_transition_full}
		\intertext{If the ripple is empty ($\r_{u}=0$) an inactivation takes place. In this case the number of inactivations increases by one yielding}
		\Pr\{&\S{u-1}=( \c_u - \b_{u},\b_{u}, \nu+1) | \S{u}=(\c_u,0,\nu)\}  \\&= \binom{\c_u}{\b_u} {\pru}^{\b_u} (1-\pru)^{\c_u-\b_u}.
		\label{eq:prob_transition_r_0_full}
	\end{align}
	The probability of ${\S{u-1} =(\c_{u-1},\r_{u-1}, \n_{u-1})}$ can be computed recursively via \eqref{eq:prob_transition_full}, \eqref{eq:prob_transition_r_0_full} starting with the initial condition
	\begin{align}
		\Pr\{\S{k}=(\c_k,\r_k, \n_u) \} = \binom{m}{\r} \Omega_1^\r \left( 1-\Omega_1\right)^{\c_k}
	\end{align}
	for all non-negative $\c_k, \r_k$ such that $\c_k+\r_k=m$ and $\n_k=0$.
	Finally, the distribution of the number of inactivations needed to complete the decoding process is given by\footnote{From \eqref{eq:distribution} we may obtain the cumulative distribution $F_{\Y}(\y)$. The cumulative distribution of the number of inactivations has practical implications. Let us assume the fountain decoder runs on a platform with limited computational capability. For example, the decoder may be able to handle a maximum number of inactive symbols (recall that the complexity of inactivation decoding is cubic in the number of inactivations, $\y$). Suppose the maximum number of inactivations that the decoder can handle is $\y_{\mathrm{max}}$. The probability of decoding failure will be lower bounded by $1-F_{\Y}\left(\y_{\mathrm{max}}\right)$.The probability of decoding failure is actually higher than $1-F_{\Y}(\y_{\mathrm{max}})$ since the system of equations to be solved in the \acf{GE} step might be rank deficient.}
	\begin{align}
		f_{\Y}(\y) = \sum_{ \c_0} \sum_{ \r_0} \Pr\{\S{0} =(\c_0,\r_0, \y) \}.\label{eq:distribution}
	\end{align}
	In Figure~\ref{fig:mbms_dist_inact} the distribution of the number of inactivations is shown, for an \ac{LT} code with degree distribution $\Omegaone(\x)$ given in \eqref{eq:deg_dist_mbms}, input block size $k = 300$ { and relative overhead $\reloverhead = 0.02$}. The chart shows the distribution of the number of inactivations obtained through Monte Carlo simulations and by evaluating \eqref{eq:distribution}. In order to obtain the simulation results for each value of absolute overhead 10000 decoding attempts were carried out. As before, we can observe a very tight match between the analysis and the simulation results.

\section{Inactivation Decoding of Raptor Codes}\label{chap:raptor_inactivation_decoding}

The analysis introduced in Section \ref{chap:inact_analysis} holds for \ac{LT} codes. We shall see next how the proposed tools can be successfully applied to the design of Raptor codes whose outer code has a dense parity check matrix (see \cite{lazaro:JSAC}). We proceed by illustrating the impact of the \ac{LT} component of a Raptor code on the inactivation count. We then develop a methodology for the design of Raptor codes based on the results of Section \ref{chap:inact_analysis}.

\subsection{Average Number of Inactivations for Raptor Codes}
Let us now consider a Raptor code based on the concatenation of a (dense) $(h,k)$ outer code and an inner \ac{LT} code. We denote by $\hmatrixpre$ the $((h-k) \times h)$ parity-check matrix of the outer code. At the input of the Raptor encoder, we have a vector of $k$ input symbols, ${\vecu=(\Raptorinput_1,~\Raptorinput_2,~\ldots, \Raptorinput_k)}$. Out of the input symbols, the outer encoder generates a vector of $h$ intermediate symbols ${\vecv=(\Rintermsymbol_1,~\Rintermsymbol_2,~\ldots, \Rintermsymbol_h)}$.
The intermediate symbols serve as input to an \ac{LT} encoder which produces the codeword ${\mathbf{\Rosymb}=(\Rosymb_1, \Rosymb_2, \ldots, \Rosymb_n)}$, where $n$ can grow unbounded. The relation between intermediate and output symbols can be expressed as $\mathbf{\Rosymb} = \vecv \G$
where $\G$ is the generator matrix of the \ac{LT} code.
The output symbols are sent over a \ac{BEC}, at the output of which the receiver collects ${m=k+\absoverhead}$ output symbols, denoted as ${\mathbf{\Rrosymb}=(\Rrosymb_1, \Rrosymb_2, \ldots, \Rrosymb_m)}$.
The relation between the collected output symbols and the intermediate symbols can be expressed as
\begin{equation}
	\Grx^\t \vecv^\t   = \mathbf{\Rrosymb}^\t .
	\label{eq:g_v_y}
\end{equation}
where $\Grx$ is a  matrix that contains the  $m$ columns of $\G$ associated to the $m$ received output symbols.
Due to the outer code constraints, we have
\begin{equation}
	\hmatrixpre \vecv^\t   = \zeros^\t
	\label{eq:h_v_z}
\end{equation}
where $\zeros$ is the length-$(h-k)$ zero vector.
Let us now define the constraint matrix $\constmatrix$ of the Raptor code
\[
\constmatrix := \begin{bmatrix}
\hmatrixpre^\t |
\Grx
\end{bmatrix}.
\]
\ac{ML} decoding  can be carried out by solving the system
\begin{align}	
	\constmatrix^\t \vecv^\t=
	\begin{bmatrix}
		\zeros |
		\mathbf{\Rrosymb}
	\end{bmatrix}^\t.
	\label{eq:raptor_sys_eq}
\end{align}
If we compare the system of equations that need to be solved for  \ac{LT} and Raptor decoding, given respectively by \eqref{eq:ml_eq_sys} and \eqref{eq:raptor_sys_eq}, we can observed how Raptor \ac{ML} decoding is very similar to \ac{ML} decoding of an \ac{LT} code. The main difference lies in the fact that matrix $\constmatrix$ is formed, for the first $h-k$ columns, by the transpose of the outer code parity-check matrix. The high density of the outer code parity-check matrix lowers the probability that the ripple contains (some of) the $h-k$ output symbols associated to the zero vector in \eqref{eq:raptor_sys_eq} (this is especially evident at the early steps of the triangulation process). As a result, we shall expect the average number of inactivations to increase with $h-k$, for a fixed overhead \footnote{In practice, the outer codes used are not totally dense, but the rows of their parity check matrix are usually denser that the rows of $\G^\t$. This is, for example, usually the case if the outer code is a high rate LDPC code, since the check node degree increases with the rate.}.

\begin{figure}
	\begin{center}
		\centering
		\includegraphics[width=\figw]{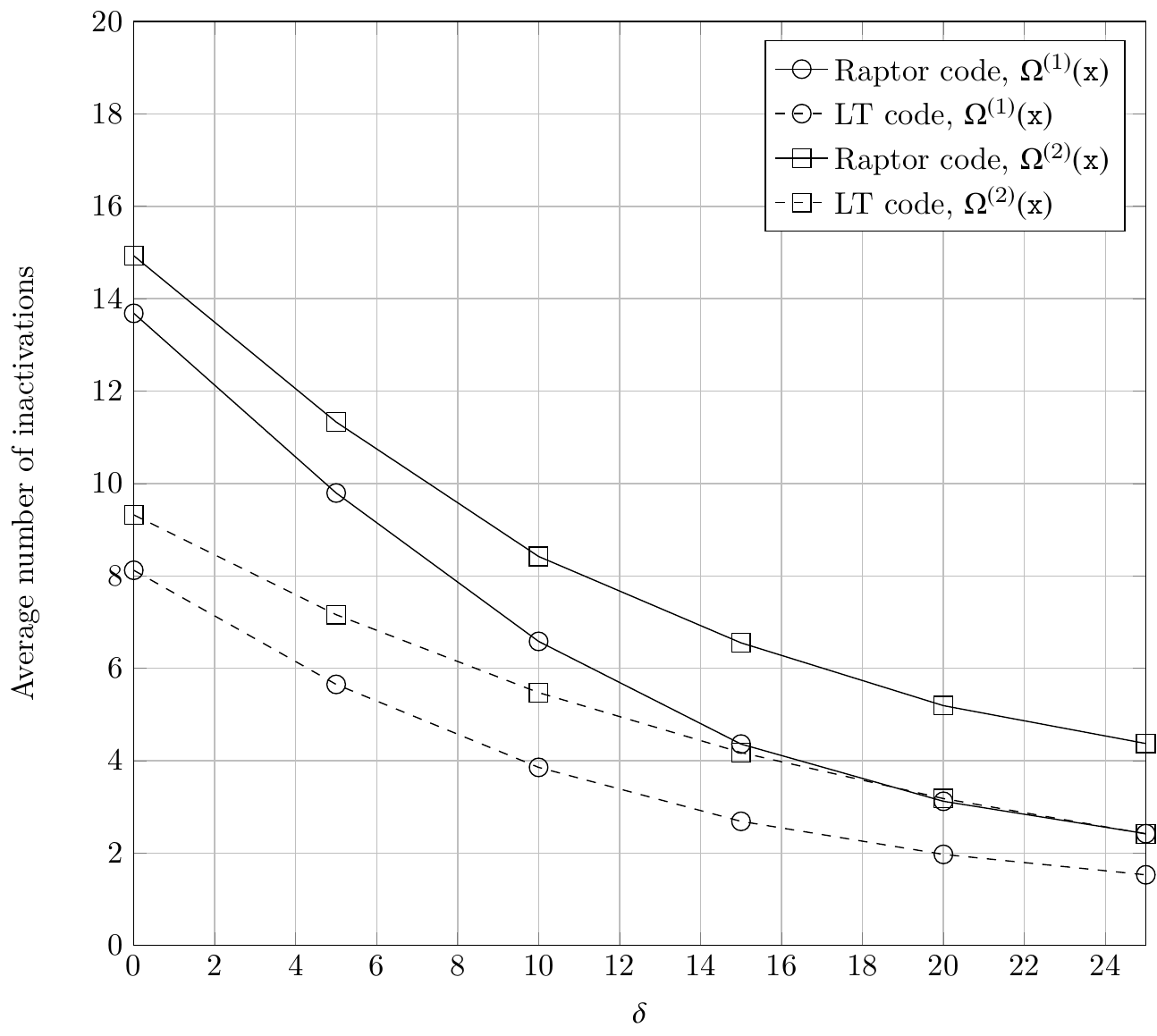}
		\caption{Expected number of inactivations two Raptor codes using a $(63,57)$ Hamming code with LT degree distributions $\Omegaone(\x)$ and $\Omegatwox(\x)$, and for two LT codes with $k=63$ and degree distributions $\Omegaone(\x)$ and $\Omegatwox(\x)$.}
		\label{fig:hamming_example}
	\end{center}
\end{figure}

In Figure \ref{fig:hamming_example} we provide the average number of inactivations needed to decode two Raptor codes, as a function of the receiver overhead $\absoverhead$. Both Raptor codes have the same outer code, a $(63,57)$ Hamming code, but different \ac{LT} degree distributions. The first distribution is $\Omegaone(\x)$ from \eqref{eq:deg_dist_mbms}, and the second distribution is
\begin{equation}
	\Omegatwox(\x):= 0.05\x + 0.2\x^2 + 0.4\x^3 + 0.3\x^4 + 0.05\x^{40}.
	\label{eq:deg_dist_example}
\end{equation}
The Figure also shows the number of inactivations needed to decode the two standalone \ac{LT} codes. If we compare the number of inactivations required by the Raptor and \ac{LT} codes, we can see how for both degree distributions, the number of additional inactivations needed for Raptor decoding with respect to \ac{LT} decoding is very similar. We hence conjecture that the impact of the (dense) outer code on the inactivation count depends mostly on the number of the outer code parity-check equations.
This empirical observation provides a hint on a practical design strategy for Raptor codes: If one aims at minimizing the number of inactivations for a Raptor code based on a given outer code, it is sufficient to design the \ac{LT} code component for a low (i.e., minimal) number of inactivations. Based on this consideration, an explicit design example is provided in the following subsection.

\subsection{Example of Raptor Code Design}\label{sec:des}
We consider a Raptor code with a $(63,57)$ outer Hamming code and we assume decoding is carried out when the absolute receiver overhead reaches $\absoverhead=\absoverhead^*$, i.e., we carry out decoding after collecting $\absoverhead^*$ output symbols in excess of $k$. Furthermore, we want to have a probability of decoding failure, $\Pf$, lower that a given value $\Pfstar$. Thus, the objective of our code design will be minimizing the number of inactivations needed for decoding while achieving a probability of decoding failure lower than $\Pfstar$. Hence, the design problem consists of finding a suitable output degree distribution for the inner \ac{LT} code. For illustration we will introduce a series of constraints in the output degree distribution. In particular we constraint the output degree distribution to have  the same maximum and average output degree as standard R10 Raptor codes ($\avgd\approx 4.63$ and $\dmax=40$, \cite{MBMS12:raptor}). Furthermore, we constraint the output degree distribution to have the same support as the degree distribution of R10 raptor codes, that is, only degrees $1,2,3,4,10,11$ and $40$ are allowed.  These design constraints allow us to perform a fair comparison between the Raptor code obtained through optimization and a Raptor code with the same outer code and the degree distribution from R10 Raptor codes.

The design of the \ac{LT} output degree distribution is  formulated as a numerical optimization problem. For the numerical optimization we used is \ac{SA} \cite{kirkpatrick1983optimization}. More concretely, we define the objective function to be minimized to be the following function \cite{lazaro:scc2015}
\[
\eta = \Exp [  \Y ] + \phi \left(\barPf\right)
\]
where $\Exp [  \Y ]$ is the number of inactivations needed to decode the \ac{LT} code and the penalty function $\phi$ is defined as
\begin{align}
	\phi \left(\barPf\right) =
	\begin{cases}
		0   & \text{if } \barPf<\Pfstar \\
		b~\left(1- \Pfstar /\barPf\right)& \mathrm{otherwise}
	\end{cases}
\end{align}
being $b$ a large positive constant\footnote{In our example, $b$ was set to $10^4$.  A large $b$ factor ensures that degree distributions which do not comply with the target probability of decoding failure are discarded.}, ${\Pfstar}$ the target probability of decoding failure at $\absoverhead=\absoverhead^*$ and $\barPf$ the  upper bound to the probability of decoding failure of the Raptor code in \cite{lazaro:Globecom2016}, which for binary Raptor codes has the expression\footnote{The use of the upper bound on the probability of decoding failure, $\barPf$ in place of the actual value of $\Pf$ in the objective functions stems from the need of having a fast (though, approximate) performance estimation to be used within the \ac{SA} recursion. The evaluation of the actual $\Pf$ presents a prohibitive complexity since it has to be obtained through Monte Carlo simulations. Note also that  the upper bound of \cite{lazaro:Globecom2016} is very tight.}
\[
\Pf  \leq \barPf := \sum_{l=1}^h \weo_{\l} \pil^{k+\absoverhead}
\]
where $\weo_l$ is the multiplicity of codewords of weight $l$ in the outer code, and $\pil$ depends on the \ac{LT} code output degree distribution as
\[
\pil = \sum_{j=1}^{\dmax} \Omega_j \sum_{\substack{i=\max(0,\l+j-h)\\ i~\mathrm{even}}}^{ \min (\l,j)} \frac{ \binom{j}{i} \binom{h-j}{\l-i} } { \binom{h}{\l}}.
\]

In particular, two code designs were carried out using the proposed optimization. In the first case we set the overhead to $\absoverhead^* = 15$ and the target probability of decoding failure to $\Pfstar=10^{-3}$, and we denote by $\Omegathree$ the distribution obtained from the optimization process. In the second case we chose the same overhead, $\absoverhead^* = 15$, and set the target probability of decoding failure to  $\Pfstar=10^{-2}$, and we denote by $\Omegafour$ the resulting distribution. The degree distributions obtained are the following
\begin{align}
	\Omegathree(\x) &= 0.0347 ~\x^{1} + 0.3338 ~\x^{2} +  0.2268 ~\x^{3} +  0.1548 ~\x^{4}  \\
	&+  0.1515 ~\x^{10} + 0.0973 ~\x^{11}  +  0.0011 ~\x^{40}
\end{align}
\begin{align}
	\Omegafour(\x) &= 0.0823 ~\x^{1} + 0.4141 ~\x^{2} +  0.1957 ~\x^{3} +  0.1272 ~\x^{4}  \\
	&+  0.0797 ~\x^{10} + 0.0762 ~\x^{11}  +  0.0248 ~\x^{40}.
\end{align}

Monte Carlo simulations were carried out in order to assess the performance of the two Raptor codes obtained as result of the optimization process. In order to have a benchmark for comparison, a third Raptor code was considered, employing the same outer code (Hamming) and the degree distribution of standard R10 Raptor codes given in \eqref{eq:deg_dist_mbms}. Note that in all three cases we consider the same outer code, a $(63,57)$ Hamming code, and thus, the number of input symbols is $k=57$. To derive the probability of decoding failure for each overhead value $\absoverhead$ simulations were run until $200$ errors were collected, whereas in order to obtain the average number of inactivations, $1000$ transmissions were simulated for each overhead value $\absoverhead$.

\begin{figure}
	\begin{center}
		\includegraphics[width=\figw]{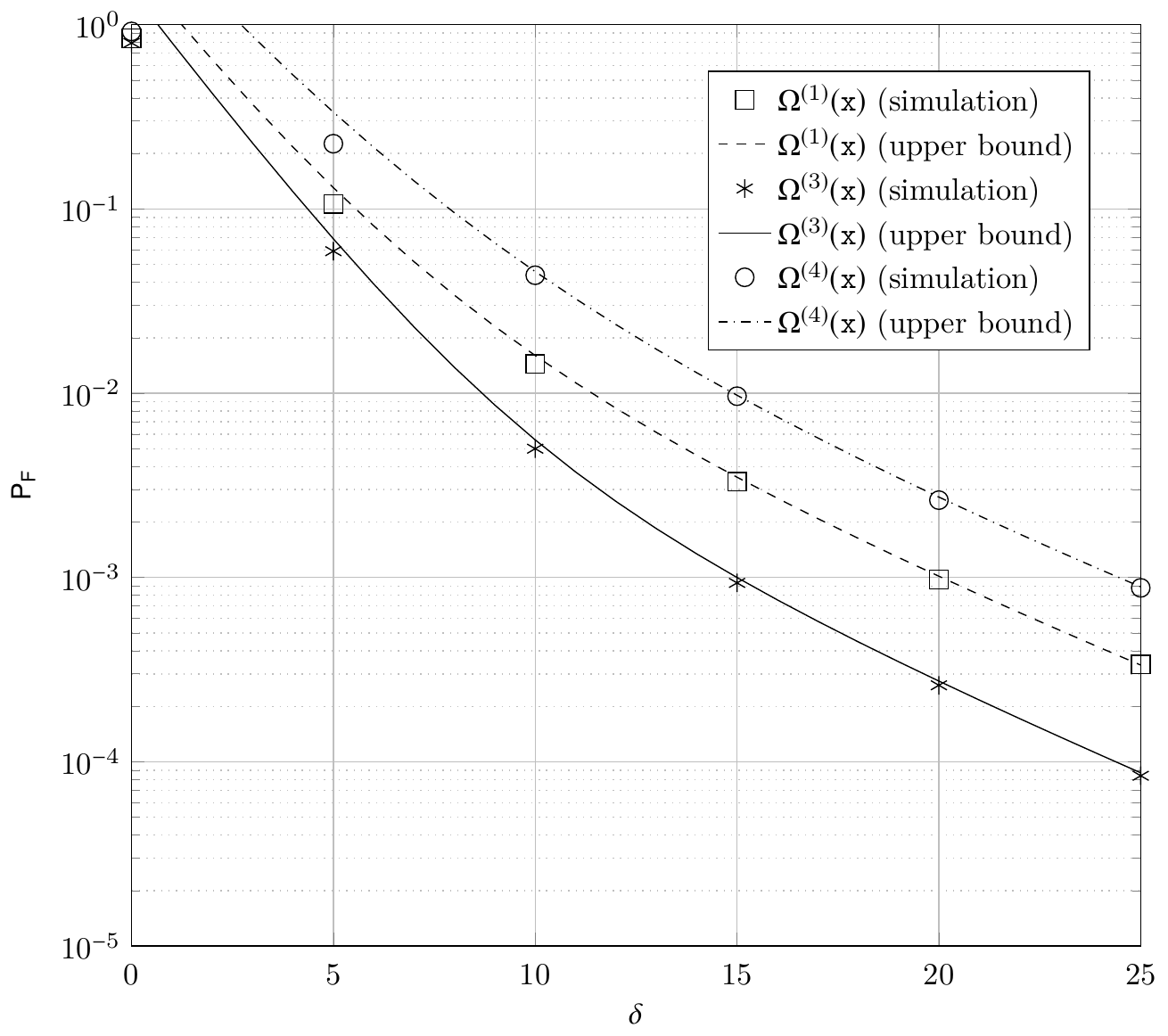}
		\centering
		\caption{Probability of decoding failure $\Pf$ vs. absolute receiver overhead $\absoverhead$ for binary Raptor codes with a $(63,57)$ Hamming outer code and \acs{LT} degree distributions $\Omegaone(\x)$, $\Omegathree(\x)$ and $\Omegafour(\x)$. The markers represent the result of simulations, while the lines represent the upper bound to the probability of decoding failure in \cite{lazaro:Globecom2016}.}
		\label{fig:overhead_perf}
	\end{center}
\end{figure}

Figure~\ref{fig:overhead_perf} shows the probability of decoding failure $\Pf$ as a function of $\absoverhead$ for the three Raptor codes based on the $(63,57)$ outer Hamming code and inner \ac{LT} codes with degree distributions $\Omegaone(\x)$,  $\Omegathree(\x)$ and $\Omegafour(\x)$. The upper bound to the probability of failure $\barPf$ is also provided. It can be observed how the Raptor codes with degree distributions $\Omegathree(\x)$ and $\Omegafour(\x)$ meet the design goal, being their probability of decoding failure at $\absoverhead=15$ below $10^{-3}$ and $10^{-2}$, respectively. It can also be observed that the probability of decoding failure of the Raptor code with degree distribution $\Omegaone(\x)$ lies between that of $\Omegathree(\x)$ and $\Omegafour(\x)$.

\begin{figure}
	\begin{center}
		\includegraphics[width=\figw]{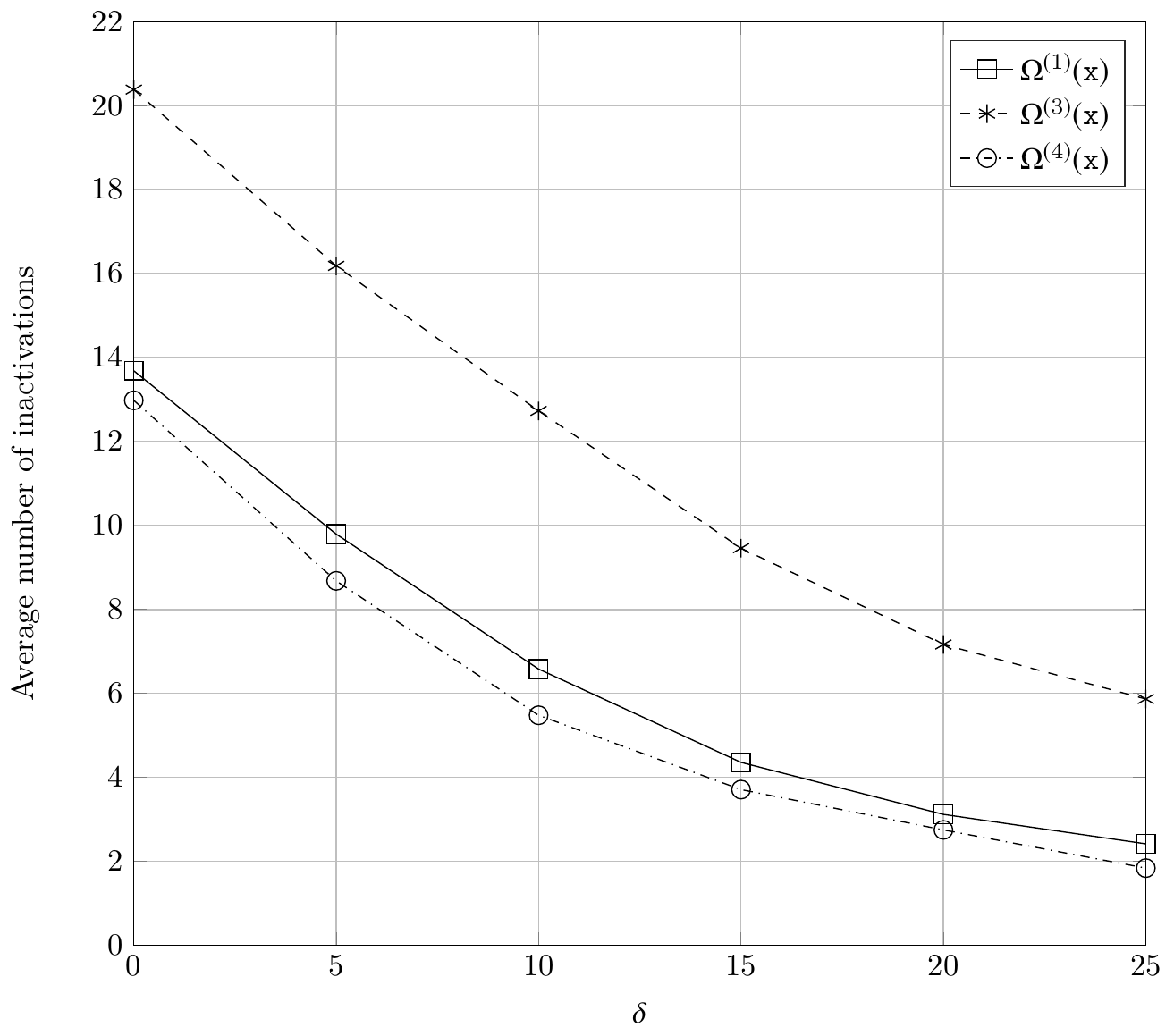}
		\centering
		\caption{Number of inactivations vs. absolute receiver overhead $\absoverhead$ for binary Raptor codes with a $(63,57)$ Hamming outer code and \acs{LT} degree distributions $\Omegaone(\x)$, $\Omegathree(\x)$ and $\Omegafour(\x)$.}
		\label{fig:inact_Hamming_Design}
	\end{center}
\end{figure}

Figure~\ref{fig:inact_Hamming_Design} shows the average number of inactivations as a function of the absolute receiver overhead for the three Raptor codes considered.  It can observed that the Raptor code requiring the least number of inactivations is $\Omegafour$, followed by $\Omegaone$, and finally the Raptor code with degree distribution $\Omegathree$ is the one requiring the most inactivations, and thus has the highest decoding complexity.

The results in Figures~\ref{fig:overhead_perf} and \ref{fig:inact_Hamming_Design} illustrate the tradeoff existing between probability of decoding failure and number of inactivations (decoding complexity): In general if one desires to improve the probability of decoding failure, it is necessary to adopt \ac{LT} codes with degree distributions that lead to a larger average number of inactivations.

\section{Conclusions}\label{sec:conclusion}
In this paper the decoding complexity of \ac{LT} and Raptor codes under inactivation decoding has been analyzed. Using a dynamic programming approach, recursions for the number of inactivations have been derived for \ac{LT} codes as a function of their degree distribution. The analysis has been extended to obtain the probability distribution of the number of inactivations.
Furthermore, the experimental observation is made that decoding a Raptor code with a dense outer code results in an increase of the number of inactivations, compared to decoding a standalone \ac{LT} code.
Based on this observation, it has been  shown how the recursive analysis of \ac{LT} codes can be used to design Raptor codes with a fine control on the number of inactivations vs. decoding failure probability trade-off.

\appendices

\section{Independent Poisson Approximation}\label{chap:inact_low_complex}

In Section~\ref{chap:inact_analysis} we have derived recursive methods that can compute the expected number of inactivations and the distribution of the number of inactivations required by inactivation decoding. The proposed recursive methods, albeit accurate, entail a non negligible computational complexity. In this appendix we propose an approximate recursive method that can provide a reasonably-accurate estimation of the number of inactivations with a much lower computational burden.

The development of the approximate analysis relies on the following definition.
\begin{mydef}[Reduced degree-$d$ set] The reduced degree-$d$ set is the set of output symbols of reduced degree $d$. We denote it by $\Rijsets{d}$.
\end{mydef}
The cardinality of $\Rijsets{d}$ is denoted by $\Rijcards{d}$ and its associated random variable by $\Rijreals{d}$.
Obviously, $\Rijsets{1}$ corresponds to the ripple. Furthermore, it is easy to see how the cloud $\cloudset$ corresponds to the union of the sets of output symbols of reduced degree higher than $1$, i.e.,
\[
\cloudset = \bigcup_{d=2}^{\dmax} \Rijsets{d}.
\]
Moreover, since the sets $\Rijsets{d}$ are disjoint, we have
\[
\Cloud = \sum_{d=2}^{\dmax} \Rijreals{d}.
\]
We aim at approximating the evolution of the number of reduced degree $d$ output symbols, $\Rijreals{d}$, as the triangulation procedure of inactivation decoding progresses.
As it was done in Section~\ref{chap:inact_analysis}, in the following a temporal dimension shall be added through subscript $u$ (recall that the subscript $u$ corresponds to the number of active input symbols in the graph). At the beginning of the triangulation process we have $u=k$, and the counter $u$ decreases by one in each step of triangulation. Triangulation ends when  $u=0$.
It follows that $\Rijset{u}{d}$ is the set of reduced degree $d$ output symbols when $u$ input symbols are still active. Moreover, $\Rijreal{u}{d}$ and $\Rijcard{u}{d}$ are respectively the random variable associated to the number of reduced degree $d$ output symbols when $u$ input symbols are still active and its realization.
We model the triangulation process by means of a finite state machine with state
\[
\S{u}:=\left(   \Rijreal{u}{1}, \Rijreal{u}{2},\ldots, \Rijreal{u}{\dmax}  \right).
\]

\begin{figure}
	\begin{center}
		\includegraphics[width=0.94\columnwidth ]{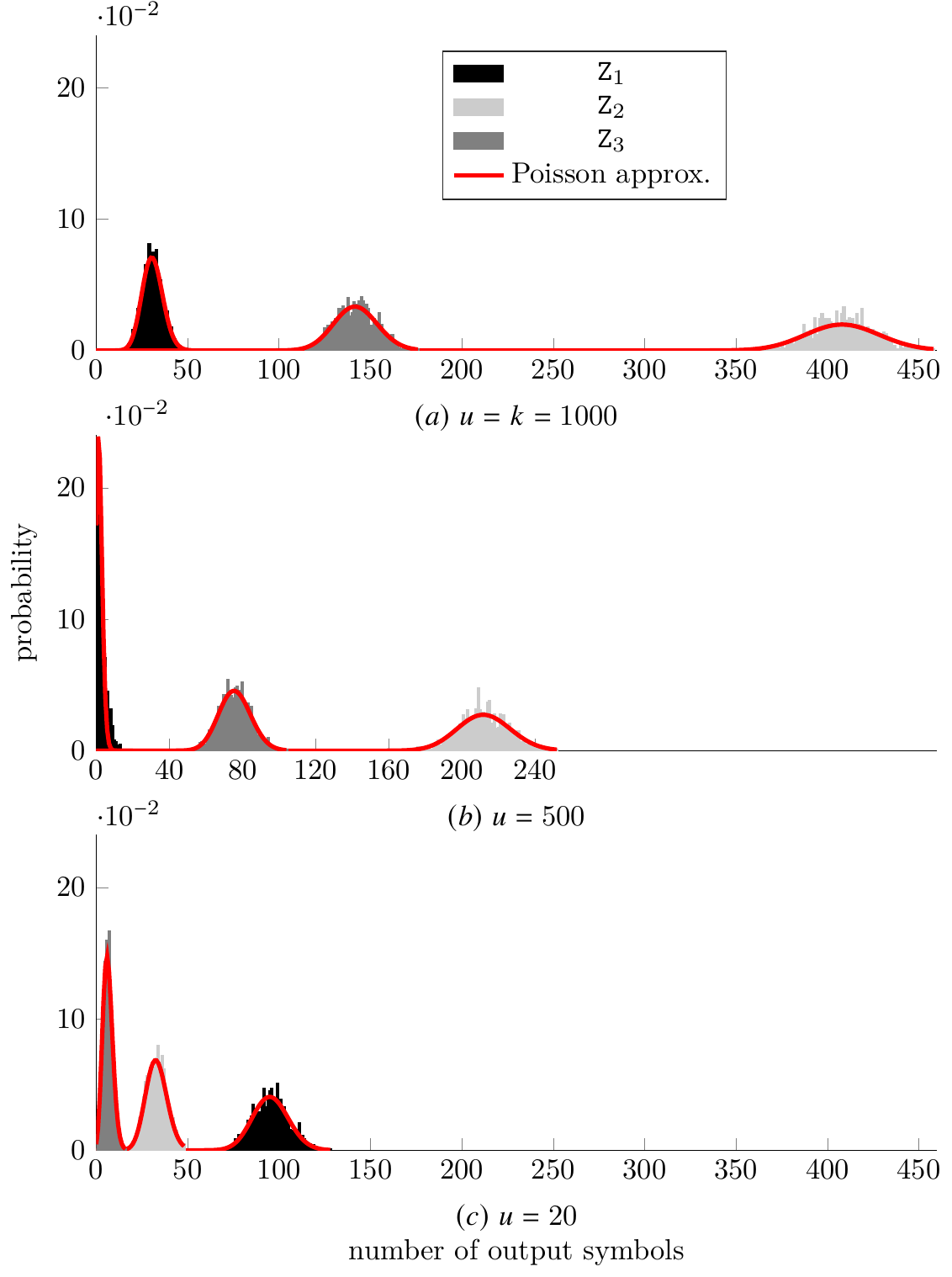}
		\centering \caption{Distribution of $\Rijreal{u}{1}$, $\Rijreal{u}{2}$ and $\Rijreal{u}{3}$ for an \ac{LT} code with \acl{RSD} and $k=1000$ obtained through Monte Carlo simulation. The upper, middle and lower figures represent respectively the distribution for $u=1000$, $500$ and $20$. The black bars represent $\Rijreal{u}{1}$, the light grey bars $\Rijreal{u}{2}$ and the dark grey bars $\Rijreal{u}{3}$. The red lines represent the best Poisson distribution fit in terms of minimum mean square error.}
		\label{fig:experimental_reduced_degree}
	\end{center}
\end{figure}

This model is equivalent to the one presented in Section~\ref{chap:inact_first_order}. However, the evaluation of the state evolution is now more complex due to the large state space. Yet, the analysis can be, greatly simplified by resorting to an approximation.
{Before decoding starts (for $u=k$), due to the independence of output symbols we have that $\S{k}$ follows a multinomial distribution, which for large number of output symbols $m$ can be approximated as the product of independent Poisson distributions.}
Figure~\ref{fig:experimental_reduced_degree} shows the probability distribution of $\Rijreal{u}{1}$, $\Rijreal{u}{2}$ and $\Rijreal{u}{3}$ for an \ac{LT} code with a \acl{RSD} and $k=1000$ obtained through Monte Carlo simulation, for $u=1000$, $500$ and $20$.  The figure also shows the curves of the Poisson distributions which match best the experimental data in terms of minimum mean square error. {As we can observe, the Poisson distribution tightly matches the experimental data not only for $u=k$, but also for smaller values of $u$}. Hence, we shall approximate the distribution of the decoder state at step $u$ as a product of independent Poisson distributions,
\begin{equation}
	\Pr \{ \S{u}= \Rijcardv{u} \} \approx \mathlarger{\prod}_{d=1}^{\dmax} \frac{ {\rij{u}{d}}^{\Rijcard{u}{d}}  e^{-\rij{u}{d}} } {\Rijcard{u}{d}!}
	\label{eq:assumption}
\end{equation}
where $\Rijcardv{u}$ is the vector defined as
$\Rijcardv{u}= \left( \Rijcard{u}{1}, \Rijcard{u}{2}, \hdots \Rijcard{u}{\dmax},  \right)$,
i.e., we assume that the distribution of reduced degree $d$ output symbols when $u$ input symbols are active follows a Poisson distribution with parameter $\rij{u}{d}$. We remark that by introducing this assumption, the number of received output symbols $m$ is no longer constant but becomes a sum of Poisson random variables. In spite of this mismatch, we will later see how a good estimate of the number of inactivations can be obtained resorting to this approximation.

Next, we shall explain how the parameters  $\rij{u}{d}$ can be determined. For this purpose let us define  $\ntx{u}{d}$ as the random variable associated with the number of output symbols of reduced degree  $d$ that become of reduced degree $d-1$ in the transition from $u$ to $u-1$ output symbols. We have
\begin{align}
	\Rijreal{u-1}{d}  &=   \Rijreal{u}{d} + \ntx{u}{d+1}  - \ntx{u}{d}.
	\label{eq:equilibrium}
\end{align}
If we take the expectation at both sides we can write
\begin{align}
	\Exp \left[ \Rijreal{u-1}{d} \right] &=  \Exp \left[  \Rijreal{u}{d}\right] + \Exp \left[ \ntx{u}{d+1}\right]  - \Exp \left[ \ntx{u}{d}\right].
	\label{eq:equilibrium_exp}
\end{align}
Let us now derive the expression of $\Exp \left[ \ntx{u}{d}\right]$. We shall distinguish two cases. First  we shall consider output symbols with reduced degree $d \geq 2$.
Since output symbols select their neighbors uniformly at random, we have that $\ntx{u}{d}$ ,  $d \geq 2$, conditioned on ${\Rijreal{u}{d} = \Rijcard{u}{d}}$ is binomially distributed with parameters $\Rijcard{u}{d}$ and $\probtx{u}{d}$, where $\probtx{u}{d}$ is  the probability that the degree of  $\rosymbrv$ decreases to $d-1$ in the transition from $u$ to $u-1$, i.e.,
\begin{align}
	\probtx{u}{d} := \Pr \{ \rosymbrv \in \Rijset{u-1}{d-1} | \rosymbrv \in \Rijset{u}{d} \}.
\end{align}
The following proposition holds.
\begin{prop}
	\label{prop:approx}
	The probability that a randomly chosen output symbol $\rosymbrv$, with reduced degree $d\geq 2$ when $u$ input symbols are active, has reduced degree $d-1$ when $u-1$ input symbols are active is
	\begin{align}
		\probtx{u}{d} = \frac{d}{u}.
	\end{align}
\end{prop}
\begin{IEEEproof}
	Before the transition, $\rosymbrv$ has exactly $d$ neighbors among the $u$ active input symbols. In the transition from $u$ to $u-1$ active symbols, $1$ input symbol is selected at random and marked as either resolvable or inactive. The probability that the degree of $\rosymbrv$ gets reduced is simply the probability that one of its $d$ neighbors is marked as resolvable or inactive.
\end{IEEEproof}

\medskip

Thus, the expected value of $\ntx{u}{d}$ is
\begin{equation}
	\Exp \left[ \ntx{u}{d}\right] = \Exp \left[ \Exp \left[\ntx{u}{d} | \Rijreal{u}{d} \right] \right]= \frac{d}{u} \, \Exp \left[  \Rijreal{u}{d} \right].
	\label{eq:ntx_d_2}
\end{equation}
If we now replace \eqref{eq:ntx_d_2} in \eqref{eq:equilibrium_exp} a recursive expression is obtained for $\rij{u}{d}$, $d \geq 2$,
\begin{align}\label{eq:recursive_poisson_d_2}
	\rij{u-1}{d} &=  \rij{u}{d} + \frac{d+1}{u} \rij{u}{d+1} - \frac{d}{u} \rij{u}{d} \\
	\rij{u-1}{d} &=  \left( 1-  \frac{d}{u} \right) \rij{u}{d} + \frac{d+1}{u} \rij{u}{d+1}
\end{align}
where we have replaced $\Exp \left[  \Rijreal{u}{d} \right] = \rij{u}{d} $ according to our Poisson distribution assumption.

We shall now consider the output symbols of reduced degree 1. In particular, we are interested in $\ntx{u}{1}$, the random variable associated to the output symbols of reduced degree $d=1$ that become of reduced degree $0$ in the transition from $u$ to $u-1$ active input symbols. Two different cases need to be considered. In the first one, the ripple is not empty, and hence there are one or more output symbols of reduced degree $1$. In this case, an output symbol $\rosymbrv$ is chosen at random from the ripple and its only neighbor $v$ is marked as resolvable and removed from the graph. Furthermore, any other output symbol in the ripple being connected to input symbol $v$ also leaves the ripple during the transition. Hence, for $\Rijcard{u}{1}  \geq 1$ we have
\begin{align}
	\Exp \left[\ntx{u}{1} | \Rijreal{u}{1} = \Rijcard{u}{1} \geq 1 \right] &= 1 + \frac{1}{u}  \left( \Rijcard{u}{1} -1 \right) \\
	&= 1 - \frac{1}{u} + \frac{1}{u}  \Rijcard{u}{1}
\end{align}
whereas for $\Rijcard{u}{1}  =0$ we have
\begin{align}
	\Exp \left[\ntx{u}{1} | \Rijreal{u}{1} = 0 \right] = 0.
\end{align}
Hence, we have
\begin{align}
	\Exp \left[\ntx{u}{1}\right] &= \left( 1 - \frac{1}{u} \right) \, \Pr\{  \Rijreal{u}{1} \geq 1\} + \frac{1}{u}  \sum_{\Rijcard{u}{1}=1}^{m} \Rijcard{u}{1} \Pr\{  \Rijreal{u}{1} = \Rijcard{u}{1}\} \\
	&= \left( 1 - \frac{1}{u} \right) \, \left( 1 - \Pr\{  \Rijreal{u}{1} =0\} \right) +  \frac{1}{u} \Exp \left[ \Rijreal{u}{1} \right] \\
	&= \left( 1 - \frac{1}{u} \right) \left( 1 - e^{- \rij{u}{1}}\right) + \frac{1}{u} \rij{u}{1}.
	\label{eq:ntx_d_1}
\end{align}
Replacing \eqref{eq:ntx_d_1} in  \eqref{eq:equilibrium_exp} a recursive expression is obtained as,
\begin{equation}\label{eq:recursive_poisson_d_1}
	\rij{u-1}{1} =  \left( 1- \frac{1}{u} \right) \rij{u}{1} + \frac{2}{u} \rij{u}{2} - \left( 1 - \frac{1}{u} \right) \left( 1 - e^{- \rij{u}{1}}\right).
\end{equation}

The decoder state probability is obtained by setting the initial condition $\rij{k}{d}= m \, \Omega_d$
and applying the recursions in \eqref{eq:recursive_poisson_d_2} and \eqref{eq:recursive_poisson_d_1}.
Furthermore, the expected number of inactivations after the $k$ steps of triangulation, can be approximated as
\begin{align}
	\Exp \left[ \Y \right] = \sum_{u=1}^{k} \Pr \{ \Rijreal{u}{1} = 0 \} \approx  \sum_{u=1}^{k} e^{-\rij{u}{1}}.
\end{align}

\begin{figure}
	\begin{center}
		\includegraphics[width=0.94\columnwidth ]{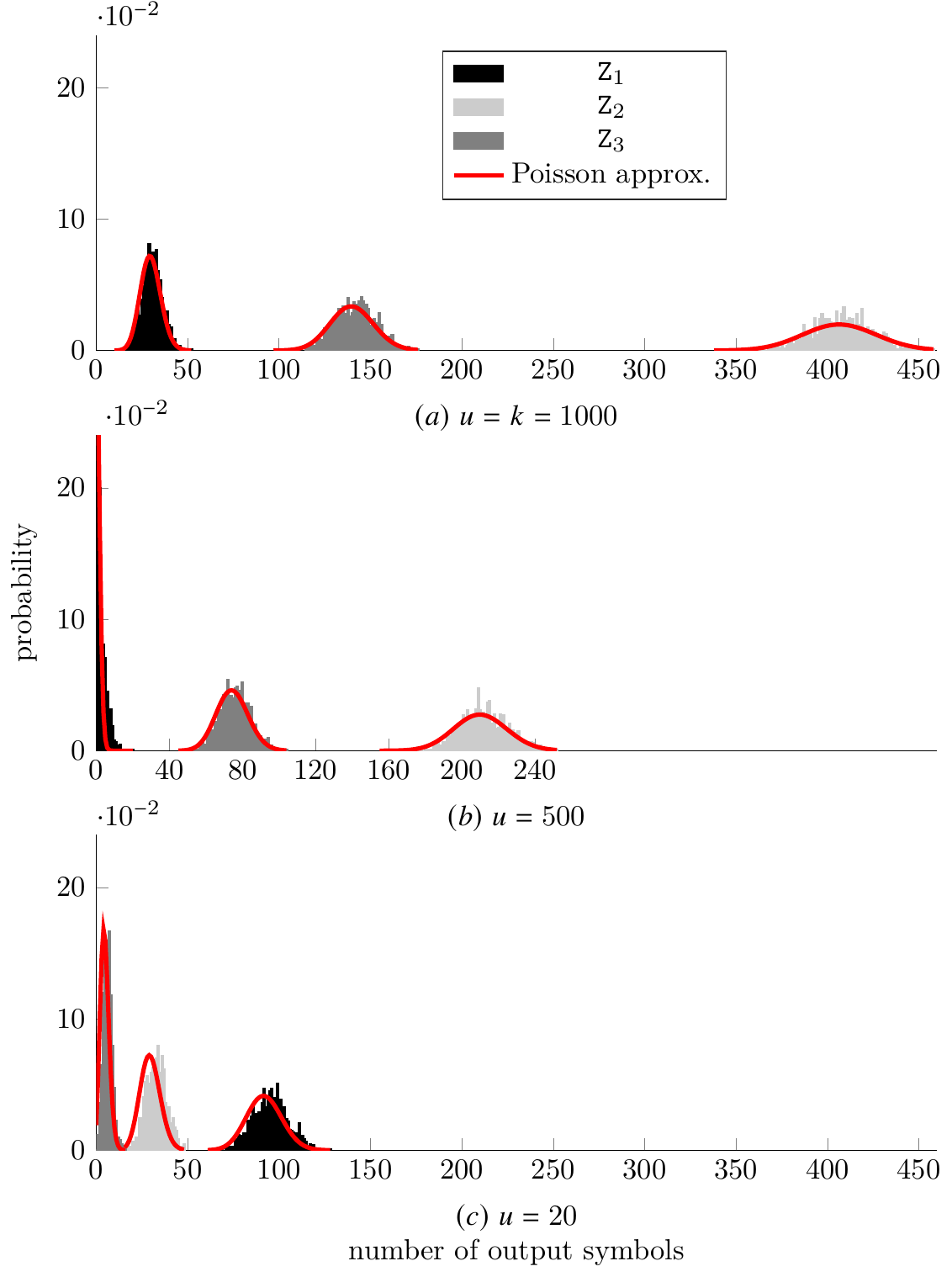}
		\centering \caption{Distribution of $\Rijreal{u}{1}$, $\Rijreal{u}{2}$ and $\Rijreal{u}{3}$ for an \ac{LT} code with \acl{RSD} and $k=1000$ obtained through Monte Carlo simulation. The upper, middle and lower figures represent respectively the distribution for $u=1000$, $500$ and $20$. The black bars represent $\Rijreal{u}{1}$, the light grey bars $\Rijreal{u}{2}$ and the dark grey bars $\Rijreal{u}{3}$. The red lines represent the Poisson distribution approximation to $\Rijreal{u}{d}$ obtained employing the model in this section.}
		\label{fig:experimental_reduced_degree_approx}
	\end{center}
\end{figure}

In Figure~\ref{fig:experimental_reduced_degree_approx} we provide again the probability distribution of $\Rijreal{u}{1}$, $\Rijreal{u}{2}$ and $\Rijreal{u}{3}$ for an \ac{LT} code with \acf{RSD} (see \cite{luby02:LT}) and $k=1000$ obtained through Monte Carlo simulation, for $u=1000$, $500$ and $20$.  The figure also shows the curves of the approximation to $\Rijreal{u}{i}$ obtained using the model in this section. We can observe how the proposed method is able to track the distribution of $\Rijreal{u}{i}$ very accurately  for $u=k$ and $u=500$. However, at the end of the triangulation process a divergence appears as it can be observed for $u=20$  in  Figure~\ref{fig:experimental_reduced_degree_approx}~(c). The source of this divergence could, in large part, be attributed to the independence assumption made in \eqref{eq:assumption}. As the number of active input symbols $u$ decreases, the dependence among the different $\Rijreal{u}{i}$ becomes stronger, and the independence assumption approximation falls apart.

\begin{figure}
	\begin{center}
		\includegraphics[width=\figw ]{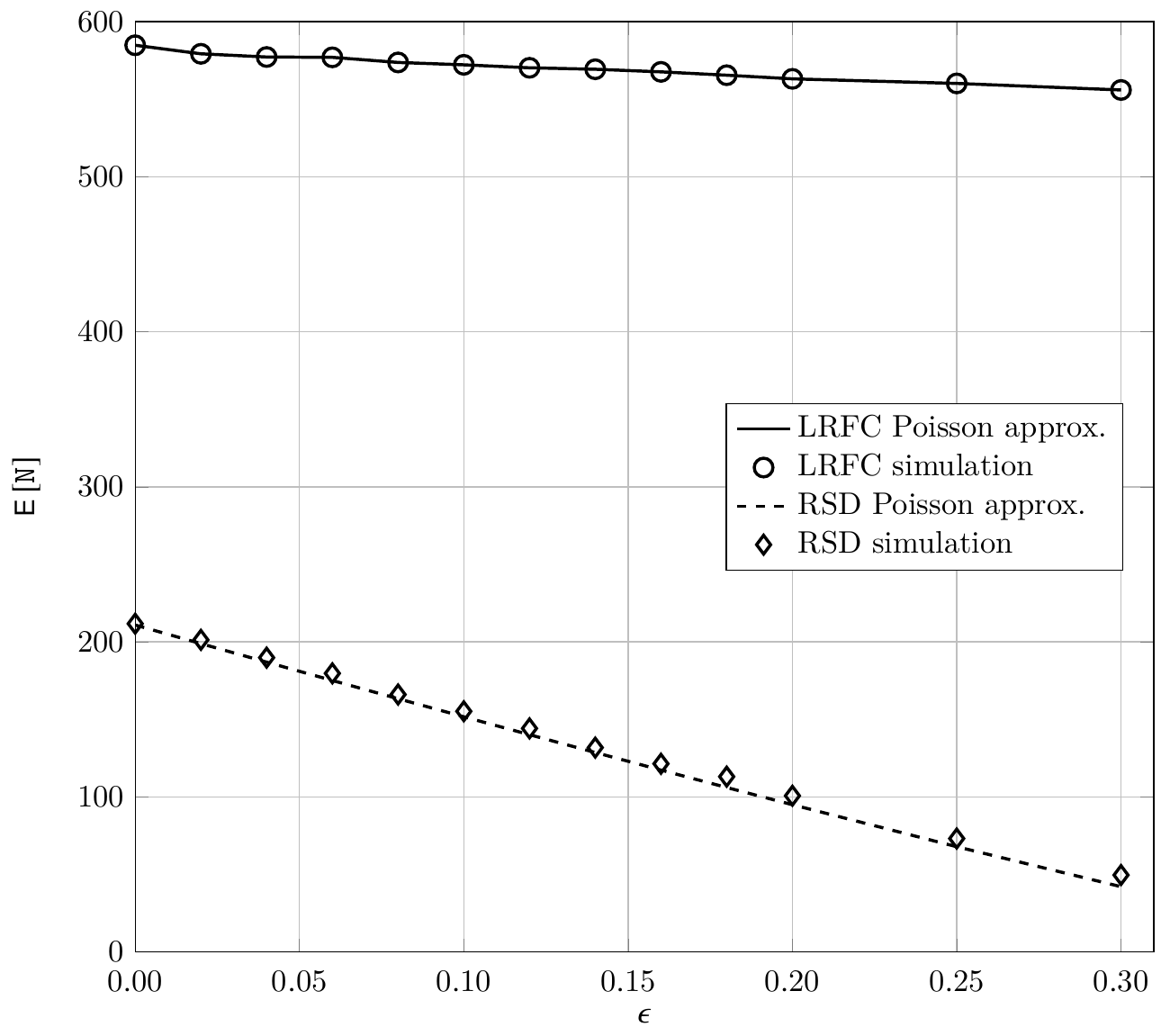}
		\centering \caption[Average number of inactivations for an \acs{LRFC} and a \acs{RSD}]{Average number of inactivations needed to decode a \acl{LRFC} and a \ac{RSD} for $k=1000$ and average output degree $\bar \Omega =12$. The markers represent simulation results and the lines represent the predicted number of inactivations using the proposed Poisson approximation.}
		\label{fig:k_1000_d_10}
	\end{center}
\end{figure}

Figure~\ref{fig:k_1000_d_10} shows the average number of inactivations needed to complete decoding for  a \ac{LRFC}\footnote{The degree distribution of a \ac{LRFC} follows a binomial distribution with parameters $k$ and $p=1/2$ (see \cite{Liva10:fountain}).	}
and a \ac{RSD}, both with average output degree $\bar \Omega =12$ and $k=1000$. The figure shows results obtained by Monte Carlo simulation and also the estimation of the number of inactivations obtained under our Poisson approximation. A tight match between simulation results and the estimation can be observed.
The experimental results indicate that, although the independence assumption made does not hold in general, it is a good approximation for most of the decoding process, deviating from simulation results only at the last stages of decoding. Thus, the proposed method can still provide a good approximation of the number of inactivations needed for decoding.

\end{document}